\documentclass[
  journal=pasa,
  manuscript=research-paper,
  year=202X,
  volume=YY,
]{cup-journal}

\usepackage{longtable}
\usepackage{amsmath}
\usepackage{amssymb,microtype,siunitx,booktabs}
\title{The initial conditions and initial mass functions of Alpha Persei, Pleiades and Praesepe}

\author{L. Hobart}
\affiliation{School of Mathematics and Physics, University of Queensland, St. Lucia, QLD 4072, Australia}
\email[L. Hobart]{l.hobart@uq.edu.au}

\author{H. Baumgardt}
\affiliation{School of Mathematics and Physics, University of Queensland, St. Lucia, QLD 4072, Australia}

\author{S. Sweet}
\affiliation{School of Mathematics and Physics, University of Queensland, St. Lucia, QLD 4072, Australia}

\received {dd Mmm YYYY}
\revised  {dd Mmm YYYY}
\accepted {dd Mmm YYYY}
\published{22 September 202X}


\begin{document}

\begin{abstract}
We have determined the initial mass function (IMF) and structural properties of the open clusters Alpha Persei, Pleiades and Praesepe using \textit{Gaia} DR3 proper motions, parallaxes and photometry. Cluster members were identified using primarily \textit{Gaia} astrometric observations, supplemented with near-infrared UKIDSS and optical HIPPARCOS survey data with stellar masses down to $0.10-0.17\,\mathrm{M}_\odot$. We measure and correct for unresolved binaries in each cluster using photometry and Monte Carlo simulations in a Bayesian framework, finding present-day unresolved binary fractions between $(20.0\pm0.8)\%$ and $(23.8\pm1.2)\%$. Through a novel approach that combines \textit{N}-body simulations with machine learning emulators and Hamiltonian Monte Carlo algorithms, we have also determined the most probable initial number of stars, binary fraction, half-mass radius and mass function under the assumption that early gas removal does not significantly influence the subsequent cluster evolution. We find a best-fitting initial mass function described by a three-stage broken power-law distribution with break masses in the ranges $0.24$–$0.50\,\mathrm{M}_\odot$ and $0.91$–$1.20\,\mathrm{M}_\odot$ and average slopes of $\alpha_\mathrm{med}=1.72\pm0.09$ and $\alpha_\mathrm{high}=2.98\pm0.22$. While the low-mass slope remains sensitive to the adopted mass–luminosity relation, this IMF reproduces the present-day clusters well once their dynamical evolution is taken into account. We find evidence of some scatter in the high mass IMF between the individual clusters, as described by the measured dispersions in mass function slopes of $\sigma_\mathrm{high}=0.29\pm0.16$, whereas the intermediate mass slope shows no significant variation with $\sigma_\mathrm{med}=0.00\pm0.14$. Our findings provide additional evidence for a (compared to a Salpeter MF) top-light IMF and a cluster-to-cluster variation of the IMF. They therefore provide constraints on the universality of the IMF and its dependence on environmental conditions.
\end{abstract}

\section{Introduction}

The stellar initial mass function (IMF) is an important fundamental concept in astrophysics, describing the relative distribution of stellar masses at birth within a given population. It plays a central role in connecting star formation on small scales to the evolution of galaxies, chemical enrichment, compact remnant abundance rates, and the baryonic mass budget of galaxies across cosmic time \citep[e.g.][]{bastian_universal_2010,oswalt_stellar_2013,hopkins_dawes_2018}. It also underpins star formation theory, as the result of molecular cloud contraction, fragmentation and protostar formation \citep[e.g.][]{bastian_universal_2010,krumholz_notes_2015, kroupa_initial_2026}. 

Since its first quantitative formulation by \citet{salpeter_luminosity_1955}, who found that the IMF of stars in the solar neighbourhood between 0.4 M$_\odot$ and 10 M$_\odot$ follows a power-law with slope $\alpha \approx 2.35$, the IMF has been extensively studied both theoretically and observationally. Most subsequent studies have shown that the IMF for lower mass stars in the galactic disk appears to increase less strongly towards lower masses or even flatten, but the exact form or values to describe this function are still under debate. Previous works retain the power law form of the IMF, but find various slope values for masses below 1 M$_\odot$ between $\alpha = 0.4 - 1.3$ \citep{miller_initial_1979, scalo_imf_1998, kroupa_variation_2001}. In addition, different mathematical forms have been suggested such as a log-normal mass function for stars below 1 M$_\odot$ \citep[e.g.][]{miller_initial_1979, chabrier_galactic_2003}. 

It is also currently unclear on the exact dependence of the IMF on environmental star forming conditions. Theoretical models \citep[e.g.][]{bate_origin_2005, marks_evidence_2012, tanvir_metallicity_2023} predict that the IMF may vary with metallicity and the density of the star-forming environment. \citet{tanvir_metallicity_2023} find that while higher metallicities can suppress high mass and very low mass star formation due to protostellar outflow feedback inhibiting accretion and radiation feedback suppressing fragmentation, these factors are primarily driven by variations in surface density. For some time, observations of nearby star-forming regions generally suggested a universal "canonical" IMF shape, often described by a multi-part power law or log-normal distribution at low masses \citep{kroupa_variation_2001,chabrier_galactic_2003}. However, there has been increasing evidence for variations in the IMF from these canonical values, particularly in the lower mass regime with $m<1\text{M}_\odot$. Studies of SMC and Milky Way field stars show these populations appear to be more deficient in low mass stars \citep{sollima_stellar_2019,kalirai_ultra-deep_2013,alcazar-julia_joint_2025,legnardi_small_2025}. This trend has also been observed in globular clusters \citep{baumgardt_evidence_2023} which has important consequences, particularly in regard to  compact remnant fractions. A bottom-light IMF, in comparison to the canonical IMF, would imply a larger population of more massive stars relative to the number of low mass stars for the same given cluster mass, and thus a larger population of black holes and neutron stars. In contrast, integrated-light studies of giant elliptical galaxies have found bottom-heavy IMFs dominated by low-mass stars \citep{van_dokkum_substantial_2010, conroy_counting_2012}. 

Theoretically, a number of mechanisms have been proposed to explain the shape and potential variability of the IMF, including turbulent fragmentation, competitive accretion, radiative feedback, and variations in metallicity or stellar density \citep{larson_thermal_2005,hopkins_variations_2013,offner_origin_2014}. 

Star clusters provide an excellent environment for IMF studies since they offer a sample of stars with similar distance, chemical composition and age, thus removing many degeneracies that complicate stellar mass estimates. Furthermore, the different environments in which clusters form allow to determine the relationship between the conditions of the star forming environment and the IMF. 

While there are advantages to using star clusters, measuring the IMF in star clusters is complicated by several observational and physical effects. The most fundamental limitation arises from dynamical evolution, where processes such as two-body relaxation, mass segregation, and tidal stripping can preferentially remove low-mass stars, altering the present-day mass function (PDMF) from the initial distribution \citep{baumgardt_dynamical_2003,portegies_zwart_young_2010}. Unresolved binaries further complicate mass estimates, as they can mimic single, more massive stars in photometric measurements, biasing IMF slopes if not properly accounted for \citep{kroupa_effects_1991, kroupa_variation_2001, cordoni_photometric_2023}. Open clusters present additional challenges relative to dense globular clusters. Their shallower potential wells and lower stellar densities make them more susceptible to tidal disruption by the Galactic field, leading to significant mass loss on relatively short timescales \citep{lamers_analytical_2005}. These factors can mask or mimic intrinsic variations in the IMF, hence careful correction for dynamical and observational biases is essential when interpreting cluster IMF measurements. Nevertheless, since open clusters are more metal rich, less dense and younger than globular clusters, they provide a good test for the dependence of the IMF on these environmental factors. 

There are many recent works that have estimated the mass function (MF) of open clusters and compared them to the IMF of the solar neighbourhood. A number of these MF determinations, particularly for older open clusters $>100\,\mathrm{Myr}$, have found shallower slopes in the subsolar regime while retaining field-like IMF slopes at higher masses e.g. Coma Berenices \citep{kraus_stellar_2007, tang_discovery_2019} and Hyades \citep{bouvier_brown_2008, goldman_towards_2013}. This flattening is generally attributed to dynamical evolution and mass segregation, which are well documented in older systems like Praesepe \citep{khalaj_stellar_2013} as well as some younger open clusters \citep{schilbach_population_2006}. Several studies have examined how the PDMF slope relates to cluster relaxation processes, finding that in globular clusters the slope correlates strongly with the ratio of age to relaxation time \citep{sollima_global_2017,baumgardt_catalogue_2018, ebrahimi_new_2020}. Results for open clusters also show a similar trend, though with some additional scatter between individual clusters \citep[e.g.][]{ebrahimi_family_2022}.

In an effort to explore the parameter space for IMF variations further, we obtain accurate mass function measurements for three nearby open clusters and compare them to mass function determinations of globular clusters. We determine the stellar IMF of the open clusters Alpha Persei, Pleiades and Praesepe using primarily \textit{Gaia} DR3 proper motions and photometry for cluster membership determinations. In Section~\ref{sec:membership} we describe our methods for determining cluster membership, implementing astrometric and photometric filtering using multiple survey catalogues that range from observations in bright optical passbands down to near infrared surveys to ensure the completeness of cluster members over a large range of stellar masses. We determine the PDMF of each cluster using various isochrone models and an empirically derived mass-luminosity function and compare the resulting functions from each method as described in Section~\ref{sec:PDMF}. In Section~\ref{sec:binaries} we correct the observed PDMFs for the measured unresolved binary fraction using photometry and Monte Carlo simulations. Lastly, we use \textit{N}-body simulations to train machine learning models, described in Section~\ref{sec:simulations}, allowing us to find the most likely initial conditions and mass function of each cluster and account for their dynamical evolution.

\section{Membership determination}
\label{sec:membership}

To identify cluster members, we searched for an over-density in proper motion and parallax phase space. When doing this analysis, it is important to balance the goals of retaining genuine member stars and rejecting field star contaminants. Through the use of \textit{Gaia} DR3 measurements, the impact of measurement errors in introducing contaminant stars was minimised due to the high precision astrometric and photometric data available. In this work, we did not aim to recover the full extent of each cluster’s tidal tails. Instead, our selection criteria were chosen to provide a clean and well-defined sample of bound members, optimised for comparison with the simulated cluster states described in Section~\ref{sec:simulations}.

\subsection{Praesepe}
For Praesepe, the centre was adopted as $\alpha=08^\text{h}40^\text{m}13^\text{s}$, $\delta=19^\circ37'16''$ \citep{tarricq_3d_2021}. Only \textit{Gaia} DR3 sources within $15^\circ$ of this position were considered in the analysis, which corresponds to radii of $\approx36-50$ pc at the distances of the three studied clusters. These radii are much larger than the typical tidal radius of open clusters of $\approx6$ pc \citep{piskunov_tidal_2008} and thus should include all bound cluster members. Member candidates were then distinguished from background stars using their proper motions, parallaxes, and photometry.

A $\chi^2$ test was used to filter member stars based on proper motions and parallaxes. More specifically, a given star \textit{i} has a $\chi^2$ value given by
\begin{equation}
    \chi_i^2= \frac{\left(\omega_i-\bar{\omega}\right)^2}{e_{\omega,i}^2 + \sigma_\omega^2} + \frac{\left(\mu_{\alpha,i} - \bar{\mu}_\alpha\right)^2}{e_{\alpha,i}^2 + \sigma_\alpha^2} + \frac{\left(\mu_{\delta,i} - \bar{\mu}_\delta\right)^2}{e_{\delta,i}^2 + \sigma_\delta^2} 
    \label{eq:chi2}
\end{equation}
and passes the initial filtering if
\begin{equation}
    \label{eqn:chi_cond}
    \chi^2_i \leq \chi^2_{0.99,\nu=3}\approx11.345
\end{equation}
where $\omega_i$ and $\bar{\omega}$ are the parallax of star $i$ and the mean cluster parallax, $\mu_{\alpha,i}$ and $\bar{\mu}_\alpha$ are the proper motion of star $i$ and the mean cluster proper motion in right ascension, $\mu_{\delta,i}$ and $\bar{\mu}_\delta$ are the proper motion of star $i$ and the mean cluster proper motion in declination, $e_{\omega,i}$, $e_{\alpha,i}$ and $e_{\delta,i}$ are the errors of each measurement, $\sigma_\omega$ is the intrinsic spread of parallaxes due to the line of sight depth of the cluster, $\sigma_\alpha$ and $\sigma_\delta$ are the internal 1D velocity dispersions and $\nu$ is the number of degrees of freedom.

For the internal velocity dispersions, we make conservative estimates of $\sigma_\alpha=\sigma_\delta=1\,\mathrm{mas\,yr^{-1}}$ which correspond to velocities of $\approx0.9\,\mathrm{km\,s^{-1}}$ at the distance of Praesepe. This value is similar to typical estimates of the intrinsic velocity dispersion in open clusters, which are generally of order $\lesssim1\,\mathrm{km\,s^{-1}}$, and was chosen to ensure that the selection encompasses both the intrinsic kinematic spread of cluster members and the contribution from measurement uncertainties in Gaia proper motions. The parallax dispersion was chosen to be $\sigma_\omega=0.25$, which corresponds to a line of sight radius of $\approx8$ pc at the distance of Praesepe, comparable to or slightly larger than the expected physical extent of the cluster. 

The membership threshold is also a conservative value, using the 99th percentile with 3 degrees of freedom of a $\chi^2$ distribution, meaning that only 1\% of true member stars are likely to be excluded due to measurement error assuming Gaussian errors. While these values will ensure nearly all member stars are detected, the chance of including contaminating background stars in the data set is also increased. Any remaining contamination can subsequently be filtered out using photometry measurements and cross referencing with additional catalogues. 

\begin{figure}
 \includegraphics[width=0.9\columnwidth]{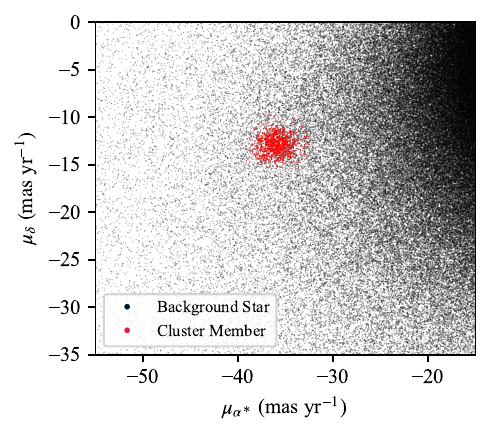}
 \caption{Proper motion distribution of stars in Praesepe in right ascension and declination. Black points show background stars whereas red points indicate cluster members identified through PM and parallax filtering. Cluster members are clearly separated from the background population, forming a compact over-density.}
 \label{fig:pm_filt}
\end{figure}

For the cluster mean parallax and proper motion, we began with the values $\bar{\omega}=5.371$ mas, $\bar{\mu}_\alpha=-36.047\,\mathrm{mas\,yr^{-1}}$ and $\bar{\mu}_\delta=-12.917\,\mathrm{mas\,yr^{-1}}$ from \textit{Gaia} DR2 \citep{gaia_collaboration_gaia_2018}. We then implemented a routine where member stars are selected using these initial reference values using the $\chi^2$ filtering. The new mean values for these three values, calculated from the resulting member population, are then used in the next iteration of the filtering on the whole population again. This process was repeated until convergence is reached and the selected member stars are consistent between successive iterations. The result of this routine can be seen in proper motion space in Figure~\ref{fig:pm_filt}, where we detect a compact over-density in proper motion and parallax out from the background stellar population.

For the photometric membership filtering of the clusters, we primarily use the G (green) and G$_\text{RP}$ (red) passbands from \textit{Gaia} DR3 \citep{gaia_collaboration_gaia_2023}. Using these two bands, we created a colour magnitude diagram (CMD) of the stars that passed the astrometric filtering criteria. Upon initial inspection of the CMD, a clear main sequence was identified. We removed a handful of stars on the upper right of the CMD that were several magnitudes from the main sequence and thus very unlikely to be true members even if they were unresolved multiple star systems. There were also a number of stars that appeared slightly below the main sequence or between the white dwarf population and main sequence. We found that most of these outliers were located spatially at the edge of the cluster and towards the edge of the cluster parameter space, with lower membership probabilities. Hence, we manually removed all of these selected stars that did not have parallax and proper motion measurements that were strongly consistent with the rest of the cluster. This ensured there was minimal contamination from non-member stars that passed the initial filtering due to coincidentally having similar astrometric properties to the cluster. 

Since we wish to measure the IMF over a broad mass range down to low mass stars, we needed to ensure the completeness of member stars detected by \textit{Gaia} at faint magnitudes. This was done by comparing identified cluster members from the process described above with those identified using a separate near-infrared survey catalogue. Due to the availability of proper motions, we use observations from the Galactic Clusters Survey (GCS) component of the UKIRT Infrared Deep Sky Survey (UKIDSS) \citep{lawrence_ukirt_2007}. Photometric observations were obtained with the Wide Field Camera \citep{casali_ukirt_2007} using the five infrared filters Z, Y, J, H and K as described in \cite{hodgkin_ukirt_2009}. All data was taken from the latest data release UKIDSSDR11PLUS which provides ZYJHK measurements of Praesepe, Pleiades and Alpha Persei with coverages of 28, 79 and 50 sq. degrees respectively. 

\begin{figure*}
 \includegraphics[width=0.75\columnwidth]{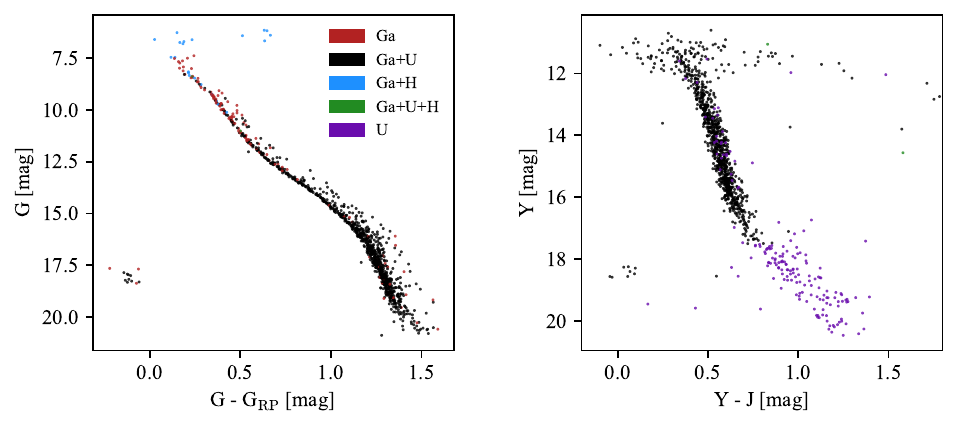}
 \caption{Colour magnitude diagrams of filtered Praesepe member stars based on the combined photometric information of \textit{Gaia} DR3, HIPPARCOS and UKIDSS. Left: Colour magnitude diagram using \textit{Gaia} G magnitudes and G-G$_\text{RP}$ colours. Right: Colour magnitude diagram using UKIDSS J magnitudes and Y-J colours. The colour coding of each star indicates which catalogue it is observed in: either Ga (\textit{Gaia}), U (UKIDSS), H (HIPPARCOS) or a combination of the three.}
 \label{fig:CMD_prae}
\end{figure*}

When finding cluster member stars from the UKIDSS catalogue, we used a method adapted and slightly changed from \cite{deacon_proper_2004} due to the much larger uncertainties in proper motion measurements in the UKIDSS data, compared to \textit{Gaia}. The method, described in Appendix~A of \cite{deacon_proper_2004}, is a maximum likelihood method which fits cluster stars with a circularly symmetric Gaussian distribution in the proper motion phase space, and field stars with a decaying exponential in the direction of the cluster's motion and Gaussian perpendicular to it. Initially, stars were filtered using CMDs constructed using photometry in Z-Y and Y-J passbands where the main sequences could be identified using the member stars from the Gaia analysis, cross referenced between catalogues using sky positions. The maximum likelihood was then solved on this subset of stars using a numerical bisection algorithm which can then be converted to give a membership probability of each star as shown in \cite{deacon_proper_2004}. A star was considered a cluster member if its membership probability was larger than 0.75. 

In addition to cross referencing infrared surveys to ensure completeness for low mass stars, we also check the completeness of \textit{Gaia} for high mass members by comparing observations from its predecessor survey HIPPARCOS \citep{esa_hipparcos_1997}. This serves as a useful comparison as the detectors and optical passbands of HIPPARCOS were able to measure stars with magnitudes as bright as $m_V=-1.44$ without saturating, meaning we can check for and include any bright cluster stars that are beyond the upper limit in brightness of \textit{Gaia}. The HIPPARCOS catalogue contains proper motions and parallaxes over full sky observations, similar to \textit{Gaia}, although with much larger uncertainties. We filtered this data similarly to the process described for the \textit{Gaia} data using the same $\chi^2$ condition. Due to the fewer observations, and thus smaller sample statistics, the mean values used in Equation~\eqref{eq:chi2} were found using the HIPPARCOS observations that were matched to the filtered \textit{Gaia} subset on their sky positions. We then used the photometry in the V and B bands to confirm that all identified cluster members fit along a main sequence in the CMD. 

When combining the three catalogues, we prioritise any \textit{Gaia} measurements and membership probabilities due to their significantly higher accuracy when a star is measured in multiple catalogues. More specifically, if a star was identified as a member in either UKIDSS or HIPPARCOS but the same star was rejected based on its \textit{Gaia} data, it was still rejected from the final dataset and if a star was accepted based on its \textit{Gaia} data, it was accepted regardless of the membership results from the other catalogues. 

For Praesepe, we find a total of 1274 member stars. Of these stars, 938 were observed in both \textit{Gaia} and UKIDSS, 136 only in \textit{Gaia}, 175 faint stars only in UKIDSS, 23 in both \textit{Gaia} and HIPPARCOS and 2 stars observed in all three surveys. The magnitudes covered by the different surveys are shown in Figure \ref{fig:CMD_prae}. We observe that \textit{Gaia} includes the brightest stars in the cluster as well as faint stars all the way down to $Y\approx17$ mag. There are a handful of stars between $Y=12-17$ mag which are only identified in UKIDSS and not \textit{Gaia}, although there are many more at brighter magnitudes than $G\approx13$ which are only observed in \textit{Gaia}. This highlights that \textit{Gaia} is highly complete for the brighter cluster population, whereas UKIDSS provides deeper photometry that extends the detected main sequence to lower masses, particularly at magnitudes where the \textit{Gaia} sample becomes incomplete.

\subsection{Pleiades \& Alpha Persei}
For the Pleiades we use the same method as for Praesepe to determine cluster members. We again use \textit{Gaia} DR3 observations within a radius of $15^\circ$ around the cluster centre of $\alpha=03^\text{h}46^\text{m}24^\text{s}$ and $\delta=24^\circ06'50''$ from \citet{tarricq_3d_2021}. For the mean proper motions and parallax of the cluster we use starting values of $\bar{\omega}=7.364$ mas, $\bar{\mu}_\alpha=19.997\,\mathrm{mas\,yr^{-1}}$ and $\bar{\mu}_\delta=-45.548\,\mathrm{mas\,yr^{-1}}$ from \citet{gaia_collaboration_gaia_2018}. We use the same estimates as Praesepe for dispersion values in the initial \textit{Gaia} filtering. 

After combining catalogues and applying photometric filtering we find a total of 1602 member stars for the Pleiades. Of these stars, 1174 are detected in both \textit{Gaia} and UKIDSS, 226 in UKIDSS only, 148 in \textit{Gaia} only, 41 in both \textit{Gaia} and HIPPARCOS, 9 in all three surveys and 4 in HIPPARCOS only. 

In the case of Alpha Persei, we slightly modify the filtering process outlined above due to the smaller absolute proper motion of the cluster, which prevents the initial $\chi^2$ filtering from reaching a stable solution for the cluster parameters due to the dense stellar background. The initial dataset was first filtered using the same maximum likelihood method as applied to the UKIDSS data in the previous two clusters. A membership probability threshold of 0.75 was set to remove any very low probability background stars from the \textit{Gaia} dataset. Since this step removed most of the background stars but still included a significant number of contaminants, we then applied the $\chi^2$ filtering from Equation~\eqref{eq:chi2} which was able to reach a stable solution. 

\textit{Gaia} data was again obtained within a $15^\circ$ radius of the cluster centre at $\alpha=03^\text{h}26^\text{m}28^\text{s}$ and $\delta=48^\circ58'30''$ from \citet{tarricq_3d_2021}. The initial proper motions and parallax were set to be $\bar{\omega}=5.718$ mas, $\bar{\mu}_\alpha=19.997\,\mathrm{mas\,yr^{-1}}$ and $\bar{\mu}_\delta=-45.548\,\mathrm{mas\,yr^{-1}}$ \citep{gaia_collaboration_gaia_2018}. 

After merging the catalogues, we found there was a halo of stars around the cluster at radii greater than 6 deg that were only detected in the \textit{Gaia} dataset. These stars were scattered randomly above and below the apparent main sequence in the cluster CMD and were thus considered non-members and removed. After photometric filtering, we found a total of 1046 member stars for Alpha Persei. 672 of these members were found in both \textit{Gaia} and UKIDSS, 201 were only in UKIDSS, 125 were only in \textit{Gaia}, 45 were in both \textit{Gaia} and HIPPARCOS and 3 stars were in all three surveys.

\subsection{Summary of astrometric and kinematic parameters}
In Table \ref{tab:astro_summ}, we present the astrometric and kinematic parameters derived in this work from the combined datasets of each cluster. For the cluster centre coordinates $\alpha_0$ and $\delta_0$, we use the density centre method from \citet{von_hoerner_numerische_1960, von_hoerner_numerische_1963} which calculates the average position of stars weighted by the local stellar density. Our derived centres agree well within a few arcminutes to those found by \citet{gaiacollaborationmapping2023}.

For the proper motions $\mu_{\alpha,\text{c}}$ and $\mu_{\delta,\text{c}}$, radial velocity $v_\mathrm{r}$ and parallax $\omega$, we find the mean and dispersion values of each cluster with uncertainties using the maximum likelihood approach outlined in \citet{pryor_velocity_1993}, which accounts for the individual measurement errors of each star. We also report the mean distance of each cluster $d$ by simply taking the inverse of the mean parallax.

The mean distances we obtain are consistent with existing studies. For the Pleiades, we find a distance of $135.90\pm0.10$ pc confirming newer estimates such as the VLBI result of $136.2\pm1.2$ pc from \citet{melis_vlbi_2014}, and ruling out the previous HIPPARCOS estimate of $120.2\pm1.9$ pc \citep{van_leeuwen_parallaxes_2009}. The derived 1D internal velocity dispersions of $\sigma_\mathrm{v}\approx0.7-0.8\,\mathrm{km\,s^{-1}}$ are typical for nearby open clusters and consistent with earlier findings \citep[e.g.][]{tarricq_3d_2021}. The mean radial velocities are also in good agreement with \citet{tarricq_3d_2021} except for Alpha Persei, which they found to have a low, yet positive value of $0.72\pm0.16\,\mathrm{km\,s^{-1}}$. From our analysis, we find the cluster to have a slightly negative radial velocity of $-0.76\pm0.27\,\mathrm{km\,s^{-1}}$, about $1\,\mathrm{km\,s^{-1}}$ different from this previous estimate. 

\begin{table}
    \centering
    \begin{tabular}{lccc}
        Parameter & Praesepe & Pleiades & Alpha Persei \\
        \hline
        $\alpha_0$ & $08^\text{h}40^\text{m}12^\text{s}$ & $03^\text{h}46^\text{m}47^\text{s}$ & $03^\text{h}26^\text{m}15^\text{s}$ \\
        $\delta_0$ & $+19^\circ38'55''$ & $+24^\circ08'13''$ & $+48^\circ54'50''$ \\
        $\mu_{\alpha,\text{c}}$ (mas/yr) & $-31.455\pm0.034$ & $12.792\pm0.019$ & $18.941\pm0.030$ \\
        $\mu_{\delta,\text{c}}$ (mas/yr) & $-11.290\pm0.028$ & $-29.144\pm0.023$ & $-21.089\pm0.031$ \\
        $\sigma_\mathrm{v}$ (km/s) & $0.809\pm0.017$ & $0.739\pm0.011$ & $0.799\pm0.017$ \\
        $v_\mathrm{r}$ (km/s) & $34.59\pm0.19$ & $5.23\pm0.26$ & $-0.76\pm0.27$ \\
        $\omega$ (mas) & $5.4028\pm0.0043$ & $7.3584\pm0.0055$ & $5.7188\pm0.0069$ \\
        $\sigma_\omega$ (mas) & $0.110\pm0.004$ & $0.177\pm0.005$ & $0.175\pm0.006$ \\
        $d$ (pc) & $185.09\pm0.15$ & $135.90\pm0.10$ & $174.86\pm0.21$ \\
    \end{tabular}
    \caption{Astrometric and kinematic parameters derived in this work for Praesepe, Pleiades and Alpha Persei. Reported values include the calculated cluster density centre in RA ($\alpha_0$) and DEC ($\delta_0$), mean proper motions in RA ($\mu_{\alpha,\text{c}}$) and DEC ($\mu_{\delta,\text{c}}$), internal 1D velocity dispersion ($\sigma_\mathrm{v}$), mean radial velocity ($v_\mathrm{r}$), mean parallax ($\omega$) with dispersion ($\sigma_\omega$) and distance ($d$). Uncertainties correspond to 1$\sigma$ errors.}
    \label{tab:astro_summ}
\end{table}

\section{Present day mass function}
\label{sec:PDMF}

To measure the observable mass functions of each cluster, we first estimated the mass of each star by fitting theoretical isochrone models. We also investigated the effects of using various conversion models on these mass estimates. In this section we investigate two theoretical isochrone models and how they affect the mass function form and values. We compare the results of these interpolations to an empirical model that uses a collection of recent eclipsing binary star mass and magnitude measurements. These mass estimates from each model were then used to find the best fitting form and values of the PDMF for each cluster using a Bayesian model. 

\subsection{Isochrone selection}

To estimate the mass of observed cluster member stars, we used PARSEC v2.0 isochrones \citep{nguyen_parsec_2022} and Mesa Isochrones and Stellar Tracks (MIST) \citep{choi_mesa_2016}. These models were selected because they incorporate updated input physics, bolometric corrections in the \textit{Gaia} photometric system, and have parameter grids that extend to the low ages of the observed clusters, including pre-main sequence evolution tracks.

For both models, we adopted previously measured values for the age and metallicity of each cluster. For the Pleiades, we adopt an age of $\approx118\,\mathrm{Myr}$ based on lithium depletion boundary (LDB) measurements, which provide a largely model-independent and precise chronometer for young clusters \citep[e.g.][]{navascues_spectroscopy_2004, frasca_lamost_2025}. For metallicity, we adopt [M/H] $\approx+0.03$ dex from high resolution spectroscopic analyses of abundances \citep[e.g.][]{soderblom_metallicity_2009, grilo_chemical_2024, frasca_lamost_2025}. We also corrected for reddening in the Pleiades isochrone using $A_v = 0.12$ mag \citep{odell_new_1994}. 

For Alpha Persei, we adopt an age of $\approx80\,\mathrm{Myr}$, primarily determined by LDB and gyrochronology measurements \citep[e.g.][]{stauffer_keck_1999, navascues_spectroscopy_2004, galindo-guil_lithium_2022, boyle_stellar_2023}. The cluster's metallicity was taken to be [M/H] $\approx+0.02$ dex based on spectroscopic studies \citep[e.g.][]{boesgaard_lithium_1988, boyle_stellar_2023}. Similar to the Pleiades, we corrected for reddening in Alpha Persei using an extinction of $A_v = 0.30$ mag \citep{prosser_membership_1992}.

\begin{figure*}
 \includegraphics[width=0.85\textwidth]{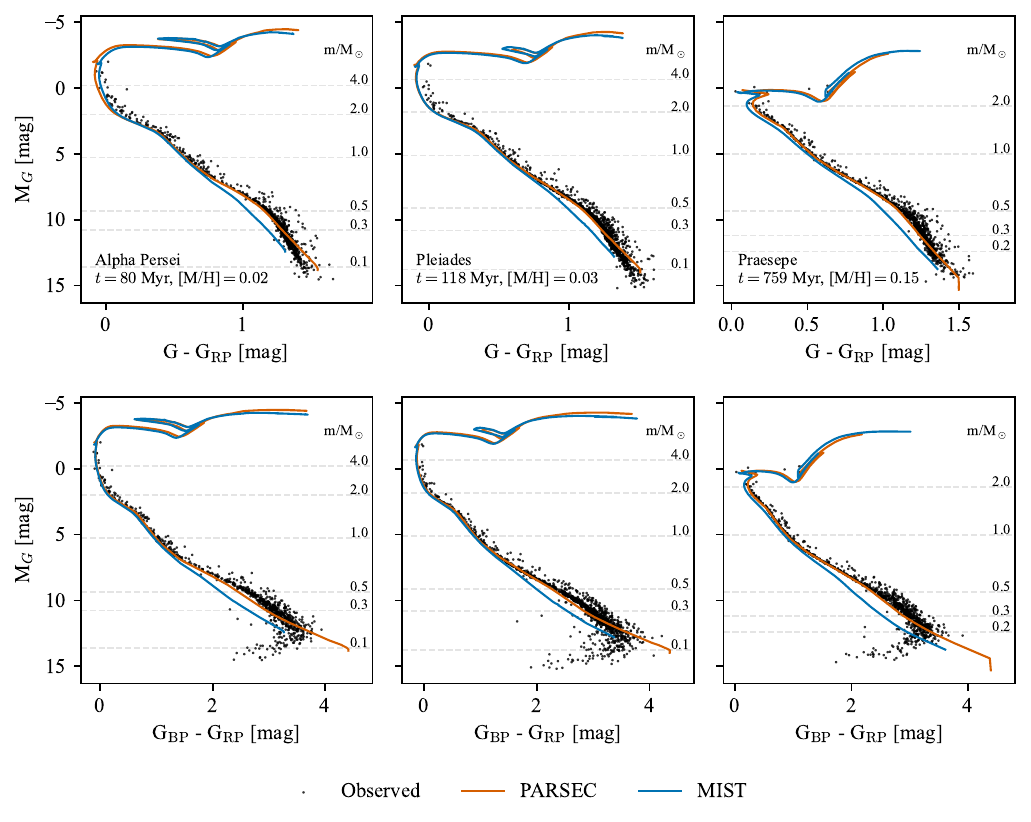}
 \caption{Colour-magnitude diagrams (CMDs) for Alpha Persei, Pleiades and Praesepe (left to right). The top row shows the absolute magnitude M$_G$ against (G-G$_\text{RP}$), while the bottom row shows M$_G$ against (G$_\text{BP}$-G$_\text{RP}$). In each panel, the PARSEC v2.0 isochrone using values from prior work is shown in orange and the corresponding MIST isochrone in blue. Grey dashed lines indicate masses along the isochrone for PARSEC models. Some The comparison illustrates the agreement between the two stellar evolution models at higher masses but some differences in photometry for stars fainter than M$_G\approx7$ mag. }
 \label{fig:Isochrones}
\end{figure*}

\begin{figure*}
    \centering
    \includegraphics[width=0.60\textwidth]{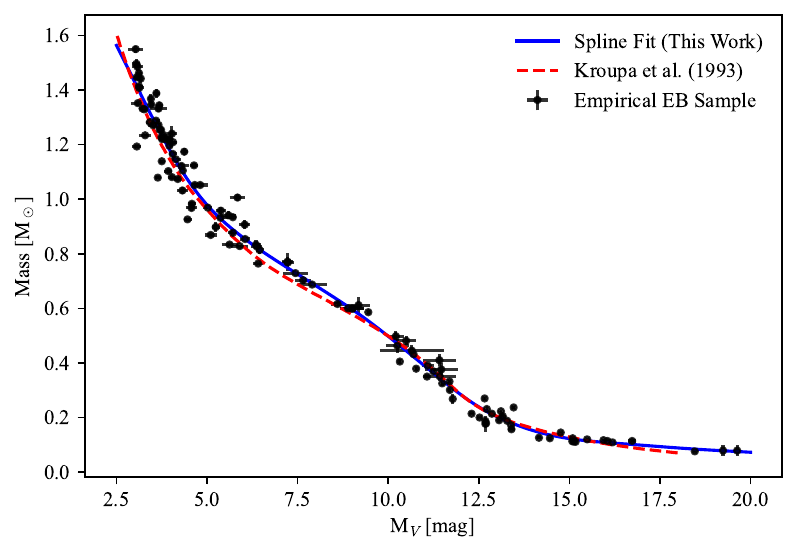}
    \caption{Empirical mass–luminosity relation for low-mass stars. The black points show the sample of eclipsing binaries used to calibrate the relation given in Table~\ref{tab:EncBinaryData}. The blue curve shows the spline fit derived in this work, while the dashed red line shows the spline fit from \citet{kroupa_distribution_1993}. The spline fit provides a smooth empirical calibration of the mass–luminosity relation across the sampled magnitude range and is adopted in this work to estimate stellar masses from M$_V$.}
    \label{fig:empirML}
\end{figure*}

For Praesepe, we adopt the parameters derived by \citet{alfonso_gaia_2023} who report a best-fitting isochrone age of $\log(\mathrm{Age}) = 8.88$ ($\approx759\,\mathrm{Myr}$) and a bulk metallicity of $Z = 0.020$ ([M/H]$\,\approx0.15$) from \textit{Gaia} photometry isochrone fitting. These values are in agreement with other recent isochrone estimates \citep[e.g.][]{brandt_age_2015, gaia_collaboration_gaia_2018, tarricq_3d_2021}. 

Figure~\ref{fig:Isochrones} shows the adopted isochrones overlaid on the cluster CMDs. At intermediate and high stellar masses, the PARSEC and MIST models are in close agreement with one another as well as the observed cluster data. At lower masses, however, we observe that the two models diverge around M$_G\approx7\,\mathrm{mag}$ ($m\approx0.7\,\mathrm{M}_\odot$). These discrepancies likely reflect differences in the models such as the treatment of convection, atmospheric boundary conditions, and low-temperature opacities. The PARSEC models better reproduce the observed photometry of each cluster at low masses, though the variation between the models indicates a degree of uncertainty, particularly for MIST models, in the theory of low mass stellar evolution used in these codes. 

Stellar masses were assigned via isochrone interpolation in three photometric bands: G, G$_\text{BP}$, and G$_\text{RP}$. For each star, we computed a $\chi^2$ statistic of the form
\begin{equation}
\chi^2(m) = \sum_{i \in \{G, G_{BP}, G_{RP}\}} 
\frac{\left(M_{i,\mathrm{obs}} - M_{i,\mathrm{iso}}(m)\right)^2}
{\sigma_{i}^2},
\end{equation}
where $M_{i,\mathrm{obs}}$ and $\sigma_i$ are the observed magnitudes and uncertainties, and $M_{i,\mathrm{iso}}(m)$ is the model magnitude at mass $m$ along the isochrone. The $\chi^2$ values were converted into relative likelihoods via
\begin{equation}
\mathcal{L}(m) \propto \exp\left(-\frac{\chi^2(m)}{2}\right),
\end{equation}
which were then normalised to obtain a posterior probability distribution function (PDF) in mass under the assumption of uniform priors. The adopted stellar mass corresponds to the median value of this PDF, and the $1\sigma$ uncertainty was defined from the central 68\% interval.

This procedure was repeated independently for the PARSEC and MIST grids, enabling a direct comparison of derived masses and an assessment of model-dependent uncertainties in the resulting PDMF.

\subsection{Empirical mass-luminosity relation using eclipsing binaries}
\label{sec:empiricalML}

As there is evidence that the theoretical mass conversion from the two investigated isochrone models, particularly for low mass stars, may be inconsistent or uncertain, we choose to estimate stellar masses using an independent, empirically determined mass-luminosity (M-L) relationship. To achieve this, we use a similar method to \citet{kroupa_distribution_1993} who fit a model to mass and magnitude measurements of eclipsing binary systems from \citet{popper_stellar_1980}. In this work, we use a collection of recent eclipsing binary measurements with a focus on low mass stars since we are more confident in the isochrone modelling for high mass stars $>1\,\mathrm{M}_\odot$. The individual star data used is tabulated in Table~\ref{tab:EncBinaryData}; the majority of these observations come from previous collections compiled in either \citet{torres_accurate_2010}, \citet{benedict_solar_2016} or \citet{iglesias-marzoa_refined_2017}. 

It is important to note that this relationship is determined using M$_V$ measurements as it has historically been the standard optical passband for eclipsing binary analyses and nearby benchmark stars, and is not directly equivalent to \textit{Gaia} $G$-band photometry. Thus, we first compute the fit between mass and M$_V$ using the binary star data and then convert the \textit{Gaia} observations of each cluster to M$_V$ using the colour-dependent transformation provided by \citet{riello_gaia_2021}, based on the Gaia EDR3 photometric calibration. 

The comprehensive catalogue presented in \citet{torres_accurate_2010} extends over a large range of stellar masses. Since we are only interested in main sequence stars, particularly with low masses, we exclude stars brighter than M$_V=3$ ($m\approx1.4 \,\mathrm{M}_\odot$) which have undergone post main-sequence evolution. We also exclude the 9 stars from this dataset with magnitudes fainter than M$_V=7.5$ ($m\approx0.7\,\mathrm{M}_\odot$) as the adopted polynomial function of bolometric corrections (BC) from \citet{flower_transformations_1996} begins to break down at very low masses. 

To better populate the parameter space with stars between $0.3-0.5\,\mathrm{M}_\odot$ we apply our own bolometric corrections to the bolometric magnitudes of six stars from \citet{birkby_discovery_2012}. To achieve this, we used observations of O9-M9 dwarf effective temperatures and BC values as determined in \citet{pecaut_intrinsic_2013} and used them to interpolate BC values based on the observed $T_{eff}$ values in \citet{birkby_discovery_2012}, with uncertainties being propagated to the final M$_V$ value.

With the data compiled, we fitted a smoothing B-spline satisfying the Generalised Cross Validation (GCV) criterion as described in \citet{wahba_spline_1990}. This relationship with respect to the eclipsing binary data and the previous fit from \citet{kroupa_distribution_1993} are displayed in Figure~\ref{fig:empirML}. We see similar behaviour between the two models, although our model does appear to fit our sample better between around $0.5-0.9\,\mathrm{M}_\odot$ and flattens more at magnitudes fainter than  M$_V\approx16$. 

While using an empirical M-L relation avoids issues related to theoretical modelling of stellar evolution, it does have several limitations in the context of star cluster mass estimates. Unlike the multi-band fitting described above, which simultaneously uses G, G$_\text{BP}$, and G$_\text{RP}$ measurements with errors, the M$_V$ relation does not account for colour information and is therefore less sensitive to subtle temperature or metallicity effects. In fact, of the low mass stars with metallicity estimates collated in \citet{benedict_solar_2016}, we find a moderately positive relationship between the residuals of our model fit and stellar metallicity with a Pearson correlation coefficient of $r=0.51$ that is statistically significant ($p=0.0023$). This would imply a star with higher metallicity may have a larger mass than one with lower metallicity and the same M$_V$. We do not, however, incorporate corrections for metallicity in this work as robust metallicity estimates are unavailable for the majority of stars in our binary sample. The potential systematic bias introduced by neglecting metallicity dependence should therefore be borne in mind when interpreting the derived mass function.

\subsection{Mass function modelling and fitting}

We infer the stellar mass function (MF) using a forward modelling approach that explicitly accounts for measurement uncertainties on individual stellar masses. We choose to model the mass function as a multistage power law function similar to \cite{kroupa_variation_2001}, for better direct comparisons with determinations of IMF values from other published work. This function is commonly broken into either two or three stages depending on the particular mass range being fitted. A three stage power law MF can be described by the function

\begin{equation}
    \xi(m) = 
    \begin{cases}
    Cm^{-\alpha_\mathrm{l}}, & m_\text{min} \leq m \leq m_{x1}\\
    Cm_{x1}^{\alpha_\mathrm{m} - \alpha_\mathrm{l}}m^{-\alpha_\mathrm{m}},  & m_{x1} \leq m \leq m_{x2}\\
    C m_{x2}^{\alpha_\mathrm{h} - \alpha_\mathrm{m}} m_{x1}^{\alpha_\mathrm{m} - \alpha_\mathrm{l}} m^{-\alpha_\mathrm{h}},  & m_{x2} \leq m \leq m_\text{max}
    \end{cases}
    \label{eqn:MF}
\end{equation}
where $C$ is a normalisation constant, $m_{x1}$ and $m_{x2}$ are the break masses between the different MF segments, $m_\text{min}$ and $m_\text{max}$ are the minimum and maximum masses, and $\alpha_\mathrm{i}$ is the mass function "slope" of mass domain i. Here, the constants $m_{x1}^{\alpha_\mathrm{m} - \alpha_\mathrm{l}}$ and $m_{x2}^{\alpha_\mathrm{h} - \alpha_\mathrm{m}} m_{x1}^{\alpha_\mathrm{m} - \alpha_\mathrm{l}}$ ensure continuity between the different power law segments. A two stage power law MF is similar in form, with only the first two segments, needing only two slope values and a single break point to be defined. 

In general, the parameter of interest is the MF slope $\alpha_i$ as it describes the relative amount of high and low mass stars in a given mass range $i$. Here we describe the fitting process in which we determined the best fitting two and three stage power law functions for each cluster in a Bayesian framework, and find the preferred model given the Bayes factor $K$. We applied this method to the stellar masses estimated using both PARSEC and MIST isochrone models as well as the empirical M-L spline fit determined in Section~\ref{sec:empiricalML} and compare the various forms and values of the PDMF inferred using each model. 

If $m_j^\text{obs}$ and $\sigma_j$ denote the observed mass and uncertainty of star $j$, then instead of treating each mass as an exact sample from a MF $\xi(m)$, we marginalise over the unobserved true mass $m_j^\text{true}$. The likelihood contribution of each observed star is then
\begin{equation}
    \mathcal{L}_j = \int_{m_\text{min}}^{m_\text{max}}\xi(m)P(m_j^\text{obs} | m,\sigma_j)dm,
\end{equation}
where $P(m_j^\text{obs} | m,\sigma_j)$ is assumed to be a Gaussian probability density centred at $m$ with dispersion $\sigma_j$. 

We compute this integral numerically over the intersection of the model bounds $\left[m_\text{min}, m_\text{max}\right]$ and a finite window $\left[m_j^\text{obs}\pm5\sigma_j\right]$ using a trapezoidal method on a fixed grid of points. The total log-likelihood is then simply given by
\begin{equation}
    \ln\mathcal{L}=\sum_{j=1}^N\ln\mathcal{L}_j.
\end{equation}

\begin{figure*}
 \includegraphics[width=0.8\textwidth]{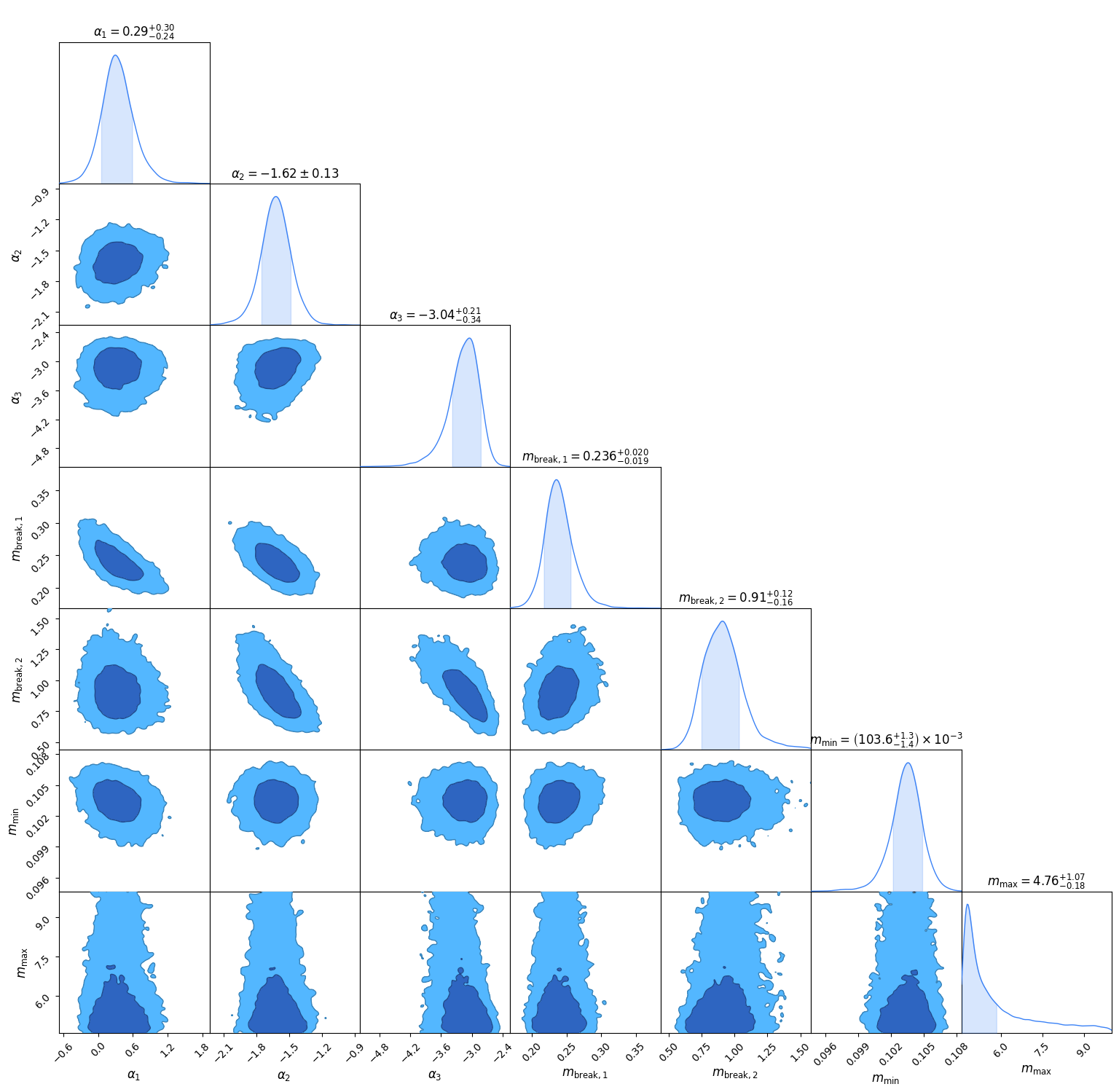}
 \caption{Posterior distributions for the parameters of the three-segment broken power-law IMF model for the Pleiades. The fitted parameters include the three power-law slopes ($\alpha_1$, $\alpha_2$, $\alpha_3$), the two break masses ($m_{\mathrm{break},1}$ and $m_{\mathrm{break},2}$), and the minimum and maximum stellar masses ($m_{\min}$ and $m_{\max}$) with $1\sigma$ credible intervals. Most parameters, excluding $m_{\max}$, are well constrained with approximately uni-modal posterior distributions, while several moderate correlations are evident between the power-law slopes and break masses.}
 \label{fig:PDMF_posterior}
\end{figure*}

This leaves us with 5 free parameters for a two stage power law fit ($\alpha_1$, $\alpha_2$, $m_x$, $m_\text{min}$, $m_\text{max}$) and 7 free parameters for a three stage power law fit ($\alpha_1$, $\alpha_2$, $\alpha_3$, $m_{x1}$, $m_{x2}$, $m_\text{min}$, $m_\text{max}$) that need to be determined. We obtain posterior distributions and the Bayesian evidence of each model using the nested sampling algorithm implemented in NAUTILUS \citep{lange_span_2023}, assuming uniform priors for each parameter. An example of a posterior distribution obtained using this method is given in Figure~\ref{fig:PDMF_posterior}, where we simultaneously fit all parameters of a three stage power law to the Pleiades masses found using the PARSEC isochrone model. We see good convergence of all parameters with approximately Gaussian errors, excluding the maximum mass bound $m_\text{max}$. 

We apply a manual lower bound to each cluster corresponding to $\mathrm{G}\approx20\,\mathrm{mag}$ for Alpha Persei and Pleiades and $\mathrm{G}\approx19.2\,\mathrm{mag}$ for Praesepe due to the completeness of \textit{Gaia} at fainter magnitudes as well as the limitations in the isochrones at any smaller masses. This corresponds to a lower limit of $0.10\,\mathrm{M}_\odot$ for Alpha Persei and Pleiades and $0.17\,\mathrm{M}_\odot$ for Praesepe. 

\begin{figure*}
 \includegraphics[width=0.85\textwidth]{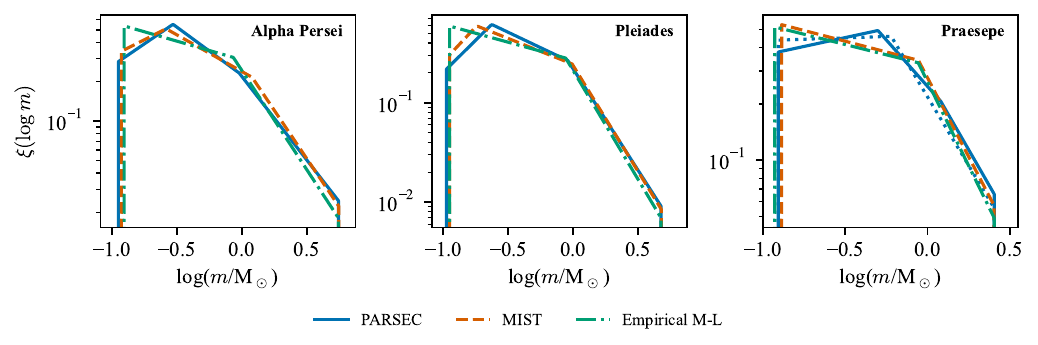}
 \caption{Present-day mass functions (PDMFs) for Alpha Persei, the Pleiades, and Praesepe (left to right). In each panel, the favoured broken power-law model (either two- or three-segment, depending on the cluster) is shown for masses estimated using three different methods: PARSEC isochrones (solid blue), MIST isochrones (dashed orange), and the empirical mass–luminosity relation (dot–dashed green). The resulting PDMF shapes are broadly consistent across the three mass estimation methods, however, the models display differences below $\approx0.3\,\mathrm{M}_\odot$. For Praesepe (right) there was no significant preference between models using PARSEC masses so both the three segment (blue line) and two segment (blue dotted) are shown.}
 \label{fig:PDMF comparison}
\end{figure*}

\begin{figure*}
 \includegraphics[width=0.9\textwidth]{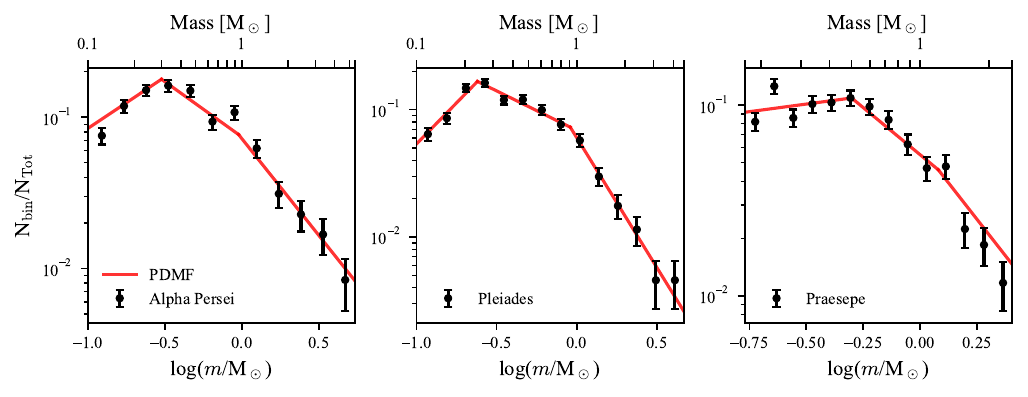}
 \caption{Cluster observed present day mass functions using PARSEC v2.0 isochrone masses, uncorrected for binaries or dynamical evolution. Black points show the fraction of stars in each log mass bin with $1\sigma$ error bars, assuming Poisson statistics. The solid red line indicates the best-fitting three-segment broken power-law model.}
 \label{fig:triple_PDMF}
\end{figure*}

When determining the optimal functional form for the PDMF, we find varying results depending on the mass estimation model used. For example, in the Pleiades we find very strong evidence ($\log K\gg2$) for a three stage fit when using either PARSEC or MIST mass estimates but find strong evidence ($\log K>1$) for a two stage fit when using the empirical M-L relation mass estimates. We find a similar trend between the three mass estimates for Alpha Persei, with the two isochrone models producing similar three stage fits and the spline fit model giving preference to a two stage fit, similar to that of the Pleiades. Praesepe, however, appears to be unique to the other clusters in that the MIST isochrone masses and the empirical M-L give evidence for a two stage PDMF with similar slope and break values. The PARSEC models, on the other hand, give no preference ($\log K<0.2$) to a 2 stage or 3 stage fit with slightly different parameter values compared to the other two fits. In this case, we choose to use the 3 stage PARSEC fit for Praesepe so that it more easily comparable to values of the other two clusters.

\begin{table}
    \centering
    \begin{tabular}{cccc}
        Parameter & Alpha Persei & Pleiades & Praesepe \\
        \hline
        $\alpha_\mathrm{h,pc}$ & $2.28^{+0.38}_{-0.13}$ & $3.04^{+0.34}_{-0.21}$ & $2.53^{+0.85}_{-0.27}$\\
        $\alpha_\mathrm{m,pc}$ & $1.73^{+0.24}_{-0.42}$ & $1.62\pm0.13$ & $1.99\pm0.47$\\
        $\alpha_\mathrm{l,pc}$ & $0.32^{+0.23}_{-0.34}$ & $-0.29^{+0.24}_{-0.30}$ & $0.84\pm0.31$\\
        $m_\mathrm{x2, pc}$ (M$_\odot$) & $0.95\pm0.34$ & $0.91^{+0.12}_{-0.16}$ & $1.20\pm0.33$\\
        $m_\mathrm{x1, pc}$ (M$_\odot$) & $0.30\pm0.06$ & $0.24\pm0.02$ & $0.50\pm0.11$\\
        \hline
        $\alpha_\mathrm{h,ML}$ & $2.52\pm0.20$ & $3.19\pm0.24$ & $2.80\pm0.40$ \\
        $\alpha_\mathrm{l,ML}$ & $1.29\pm0.08$ & $1.36\pm0.06$ & $1.22\pm0.08$ \\
        $m_\mathrm{x,ML}$ (M$_\odot$) & $0.86\pm0.08$ & $0.88\pm0.07$ & $0.88\pm0.11$ \\
    \end{tabular}
    \caption{Best-fitting present-day mass function (PDMF) parameters of Alpha Persei, Pleiades, and Praesepe using PARSEC v2.0 isochrone and empirical mass-luminosity relation mass estimates. Reported are the observed three-stage uncorrected PDMF slopes using PARSEC ($\alpha_\mathrm{h,pc}$, $\alpha_\mathrm{m,pc}$, $\alpha_\mathrm{l,pc}$) with mass break points ($m_\mathrm{x1, pc}$, $m_\mathrm{x2, pc}$) as well as the two-stage PDMF slopes using the empirical M-L masses ($\alpha_\mathrm{h,ML}$, $\alpha_\mathrm{l,ML}$) with break mass $m_\mathrm{x,ML}$.}
    \label{tab:PDMF_vals}
\end{table}

The differences between the three mass estimates for each cluster can be seen in Figure~\ref{fig:PDMF comparison}. We see that the MF slope and break mass at high masses $\approx1\,\mathrm{M}_\odot$ is consistent between the three models, giving confidence in our estimates in that mass range. This agreement begins to break down at masses $\lesssim0.5\,\mathrm{M}_\odot$, where the MF from different mass estimation models begin to diverge. Nevertheless, the broad structure of the PDMF, with a steep high-mass slope, a flattening toward lower masses, and a break near $m_x\approx1\,\mathrm{M}_\odot$ is recovered by all three approaches. The empirical two-stage fits (Table~\ref{tab:PDMF_vals}) show remarkably consistent break masses across all clusters with $m_x\approx0.86$–$0.88\,\mathrm{M}_\odot$ and low-mass slopes in the range $\alpha_\mathrm{low}\approx1.2$–$1.4$, while the high-mass slopes vary at a $\approx0.3$–$0.5$ level. These similarities suggest that the global PDMF shape is generally robust, whereas the detailed behaviour at lower masses is sensitive to the adopted mass–luminosity prescription.

Given these considerations, we chose to use the PARSEC mass estimates for the rest of the analysis of each cluster. The PARSEC models provide three stage solutions for all three clusters with statistical evidence, allowing us to be consistent in the parametrisation of the MF. Furthermore, we note that PARSEC produces better CMD fits compared to the MIST isochrones, incorporates pre-main sequence evolution and considers the age and metallicity dependence of a star's colour and luminosity. The empirical relationship, however, averages over stars with a range of metallicities and ages and does not consider pre-main sequence evolution which could be important for the low mass stars in Alpha Persei and Pleiades in particular. We therefore use PARSEC as our mass estimation, while acknowledging that systematic differences at $m \lesssim 0.5\,\mathrm{M}_\odot$ introduce model-dependent variations in the detailed low-mass behaviour. The best fitting PDMF models for PARSEC derived masses can be seen compared to the data in Figure~\ref{fig:triple_PDMF} with parameter estimations given in Table~\ref{tab:PDMF_vals}. 

We quote the observed total mass $M_\mathrm{tot}$ by simply summing all the estimated masses for each member star, and average mass $\bar{M}$ by taking the mean. The projected 2D half mass radius $r_\mathrm{h,2D}$ was found by first finding each star's radius from the cluster density centre given in Table \ref{tab:astro_summ} and then sorting the stars in order of radius to construct the cumulative mass distribution. We identify $r_\mathrm{h,2D}$ as the radius at which the cumulative is half the total mass, using linear interpolation between the two nearest points if the half mass is between two cumulative mass points. For the 3D half mass radius, we first found the heliocentric Cartesian coordinates of each member star using their right ascension, declination and parallax. We then determined the three dimensional cluster centre, weighted by stellar mass and found the radius of each cluster member. The half mass radius was then determined similarly to the 2D projected case. We estimated uncertainty values for $M_\mathrm{tot}$, $\bar{M}$, $r_\mathrm{h,2D}$ and $r_\mathrm{h,3D}$ by resampling from all cluster members and bootstrapping over 50000 iterations. The standard deviation of these measurements over all iterations describe the sensitivity of these measurements to member selection and thus are used as the quoted errors. 

The half-mass relaxation time $t_\mathrm{rh}$ of each cluster was estimated using the derivation from \citet{spitzer_dynamical_1987}, which is given by
\begin{equation}
    t_\mathrm{rh} = 0.138\frac{M_\mathrm{tot}^{1/2}r_\mathrm{h,3D}^{3/2}}{G^{1/2}\bar{M}\ln\left(\gamma\frac{M_\mathrm{tot}}{\bar{M}}\right)} \text{Myr}
\end{equation}
where $G$ is the gravitational constant and $\gamma=0.11$ \citep{giersz_statistics_1996}. Uncertainties from $M_\mathrm{tot}$, $\bar{M}$ and $r_\mathrm{h,3D}$ were also propagated to give an error estimate for each cluster. Due to its small age and large observed half mass radius, Alpha Persei is younger than one relaxation time, indicating that it has likely not been strongly affected by dynamical evolution. Pleiades and Praesepe on the other hand have ages similar to (Pleiades) or much greater (Praesepe) than their relaxation times, meaning that correcting for dynamical effects is more important when estimating the initial states of these clusters. A summary of the derived values based on estimated stellar masses is presented in Table \ref{tab:mass_features}.

\begin{table}
    \centering
    \begin{tabular}{cccc}
    Value & Alpha Persei & Pleiades & Praesepe \\
    \hline
    $M_\mathrm{tot}$ (M$_\odot$) & $518\pm21$ & $650\pm18$ & $598\pm14$\\
    $\bar{M}$ (M$_\odot$) & $0.62\pm0.03$ & $0.50\pm0.02$ & $0.574\pm0.02$\\
    $r_\mathrm{h,2D}$ (pc) & $5.93\pm0.33$ & $3.02\pm0.14$ & $3.16\pm0.15$\\
    $r_\mathrm{h,3D}$ (pc) & $6.89\pm0.29$ & $4.27\pm0.13$ & $4.92\pm0.17$\\
    $t_\mathrm{rh}$ (Myr) & $305\pm24$ & $190\pm11$ & $208\pm12$\\
    \end{tabular}
    \caption{Observed structural parameters of Alpha Persei, Pleiades, and Praesepe. Reported values include the total cluster masses ($M_\mathrm{tot}$), mean stellar masses ($\bar{M}$), projected 2D half-mass radii ($r_\mathrm{h,2D}$), 3D half-mass radius ($r_\mathrm{h,3D}$) and relaxation times ($t_\mathrm{rh}$). Uncertainties correspond to 1$\sigma$ errors.}
    \label{tab:mass_features}
\end{table}

\section{Binary fraction}
\label{sec:binaries}

So far we have assumed that all cluster members are single stars, however, as seen in the CMDs of each cluster shown in Figure~\ref{fig:Isochrones}, there are clear binary sequences above the main sequence due to the photometric contributions of unresolved secondary stars. Thus, for a true measurement of the PDMF we must correct for unresolved binary members that would change the mass function, particularly for lower mass secondary stars \citep[e.g.][]{kroupa_distribution_1993, khalaj_stellar_2013, sheikhi_binary_2016, cordoni_photometric_2023}. To achieve this, we adapt and improve upon previous Monte Carlo methods from works such as \citet{khalaj_stellar_2013} and \citet{sheikhi_binary_2016}. \citet{khalaj_stellar_2013} and \citet{sheikhi_binary_2016} use multiple Monte Carlo simulations of synthetic CMDs in order to find the best mass function and true binary fraction that reproduces the observed CMDs. These simulations aimed to both give the same observed mass function if all binary stars were measured as single systems as well as reproduce the same fraction of stick-out binaries as the real observed clusters. We use a similar method to find the optimal mass function and binary fraction which recreates the observed clusters, however we implement Monte Carlo simulations in a Bayesian approach to estimate the best-fit parameters with uncertainties, rather than a direct parameter grid search. 

\subsection{Observable binaries}
For the observed clusters, we measured the stick-out binary fraction by first modelling the main sequence below the turnoff as a high order polynomial to better match the observations of low mass stars compared to the theoretical isochrone. We then identified systems that deviated from the main sequence by 2.5$\sigma_\mathrm{p}$ or more to determine the observed binary fraction of each cluster, where $\sigma_\mathrm{p}$ is a star's photometric error in colour-magnitude space. When quoting any binary fractions, we refer to the fraction defined by 
\begin{equation}
    f_\mathrm{bin} = \frac{N_\mathrm{bin}}{N_\mathrm{sing} + N_\mathrm{bin}}
\end{equation}
where $N_\mathrm{bin}$ is the number of binary systems and $N_\mathrm{sing}$ is the number of single stars. 

\begin{figure*}
    \centering
    \includegraphics[width=\columnwidth]{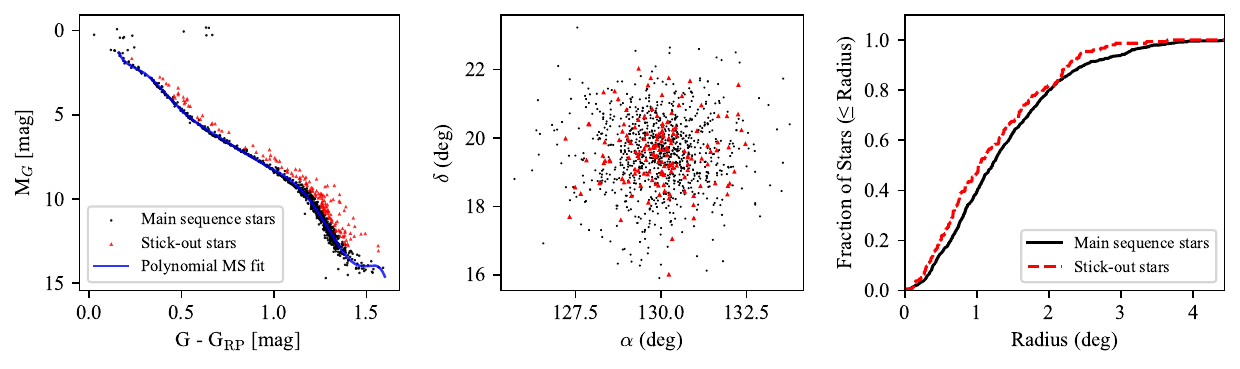}
    \caption{Identified 'stick-out' stars in Praesepe. (left) CMD of Praesepe with observable `stick-out' stars that lie above the main sequence highlighted with red triangles and the main sequence polynomial fit in blue. The CMD uses \textit{Gaia} photometry in the G and G$_\text{RP}$ passbands. (centre) Praesepe stellar spatial distribution with stick-out stars highlighted in red. (right) Cumulative fraction of stick-out stars (red dashed) and main sequence stars (black) as a function of radius. The stick-out stars are strongly concentrated to the cluster centre, indicating they are likely cluster members rather than contaminants.}
    \label{fig:Prae_Stickouts}
\end{figure*}

Figure \ref{fig:Prae_Stickouts} depicts the identified stick-out binaries in Praesepe. These systems are more concentrated toward the cluster centre, as expected for genuine binary members. Such a distribution is consistent with the effects of dynamical mass segregation, where these more massive systems would migrate inward over time. This spatial distribution therefore supports the interpretation that these stick-out stars are indeed binary systems which are cluster members, rather than contaminating stars with proper motions that coincidentally align with that of the cluster. 

Using this method, we observe the stick-out binary fraction, defined as $f_\mathrm{SO}=N_\mathrm{SO}/(N_\mathrm{MS}+N_\mathrm{SO})$, of Alpha Persei, Pleiades and Praesepe to be $(18.3\pm1.4)\%$, $(16.4\pm1.2)\%$ and $(18.8\pm1.3)\%$ respectively. The uncertainties quoted here were estimated using bootstrapping and resampling from cluster members with replacement. 

\subsection{Synthetic cluster creation}
\label{sec:synth_clusters}
To recover the true binary fraction of the cluster, we must also consider binary systems that will still lie close to the main sequence due to one star being much more massive and brighter than the other star in a given pair. We did this by creating synthetic stellar populations and CMDs in a similar procedure as in \citet{khalaj_stellar_2013}, however we implement the method in a Bayesian framework rather than a direct grid search. Thus we predict the binary corrected MF and present-day binary fraction of each cluster using the following procedure.

We first drew $N_\mathrm{sys}$ masses from a given MF, defined by three mass function slopes $\alpha_\mathrm{l}$, $\alpha_\mathrm{m}$ and $\alpha_\mathrm{h}$ and best fitting break points $m_{x1}$ and $m_{x2}$ as in Table \ref{tab:PDMF_vals} for each cluster using inverse sampling, where random uniform values are mapped through the cumulative distribution function of the MF to generate masses. The photometry of each star was then derived using the PARSEC isochrones and interpolating G, G$_\mathrm{BP}$ and G$_\mathrm{RP}$ magnitudes for each mass. Of these stars, we selected $N_\mathrm{bin}$ stars to be the primary components of binaries, where $N_\mathrm{bin}=N_\text{star obs}\cdot f_\mathrm{bin}$. For each primary component, a secondary mass was selected from a flat mass ratio distribution between $m_\mathrm{min}$ and $m_\mathrm{primary}$. The photometry of each secondary mass was derived similarly using the isochrone models and combined with the magnitudes of the primary components to give the total magnitude of each binary system as
\begin{equation}
    \text{G}_\mathrm{bin} = -2.5\log\left(10^{-0.4\text{G}_\mathrm{prim}}+10^{-0.4\text{G}_\mathrm{sec}}\right),
\end{equation}
\begin{equation}
    \text{G}_\text{BP, bin} = -2.5\log\left(10^{-0.4\text{G}_\text{BP, prim}}+10^{-0.4\text{G}_\text{BP, sec}}\right),
\end{equation}
\begin{equation}
    \text{G}_\text{RP, bin} = -2.5\log\left(10^{-0.4\text{G}_\text{RP, prim}}+10^{-0.4\text{G}_\text{RP, sec}}\right).
\end{equation}
Synthetic photometric errors were introduced using a Gaussian distribution with $\sigma_\mathrm{mag}$ equal to the median error of the photometry from the observed sample of stars. 

With this synthetic CMD, we then identified the fraction of stars that deviate significantly from the theoretical isochrone of the cluster, labelling them as stick-out binaries. We also measured the mass function slopes of the synthetic cluster with unresolved binaries so that it could be compared to the observed clusters. For a given set of tested input parameters, we created and measured 5 synthetic clusters to account for the variability in cluster generation. 

This synthetic cluster creation method was then implemented in an algorithm to estimate the Bayesian posterior and evidence and find the total binary fraction and mass function that best recreate observations of each cluster. We make use of the nested sampling algorithm in the python package NAUTILUS \citep{lange_span_2023}. Using this approach, we aim to find the combination of PDMF slopes $\alpha_\text{l, bin}$, $\alpha_\text{m, bin}$, $\alpha_\text{h, bin}$ and binary fraction $f_\mathrm{bin}$ that maximise the log likelihood given by
\begin{equation}
    \begin{split}
        \ln\mathcal{L} = -0.5 \sum_i \Bigg[ 
        \frac{(f_{\text{SO}, i} - f_{\mathrm{SO,obs}})^2}{\sigma_{\mathrm{SO,obs}}^2} 
        + \frac{(\alpha_{\mathrm{l}, i} - \alpha_\mathrm{l,\mathrm{obs}})^2}{\sigma_{\alpha_\mathrm{l}, i}^2 + \sigma_{\alpha_\mathrm{l},\mathrm{obs}}^2} \\
        + \frac{(\alpha_{\mathrm{m}, i} - \alpha_\mathrm{m,\mathrm{obs}})^2}{\sigma_{\alpha_\mathrm{m}, i}^2 + \sigma_{\alpha_\mathrm{m},\mathrm{obs}}^2} 
        + \frac{(\alpha_{\mathrm{h}, i} - \alpha_\mathrm{h,\mathrm{obs}})^2}{\sigma_{\alpha_\mathrm{h}, i}^2 + \sigma_{\alpha_\mathrm{h},\mathrm{obs}}^2}
        \Bigg]
    \end{split}
\label{eqn:bin_logl}
\end{equation}
where $i$ is the simulation trial number (1-5 here), $f_{\text{SO}, i}$ is the stick-out fraction of trial $i$, $f_{\mathrm{SO,obs}}$ is the observed cluster stick-out fraction with uncertainty $\sigma_{\mathrm{SO,obs}}$, $\alpha_{\mathrm{x}, i}$ is the apparent mass function slope of trial $i$ with uncertainty $\sigma_{\alpha_\mathrm{x}, i}$ in the mass range $x\epsilon\{\text{l,m,h}\}$ when synthetic binaries are unresolved and $\alpha_\mathrm{x,\mathrm{obs}}$ is the observed cluster mass function with uncertainty $\sigma_{\alpha_\mathrm{x},\mathrm{obs}}$ in mass range $x$. 

For computational efficiency, we measure the simulated MF in each iteration of the sampler using the maximum likelihood estimator as described in Appendix A1 from \citet{khalaj_stellar_2013}, who show that the maximum likelihood estimation (MLE) $\hat{\alpha}$ for a set of masses $x_i$ bounded by $(x_\mathrm{min}, x_\mathrm{max})$ is given by the equation
\begin{equation}
    \hat{\alpha} = 1+n\left(\sum_{i=1}^n\ln\frac{x_i}{x_\mathrm{min}}-n\frac{\ln X}{1-X^{\hat{\alpha}-1}}\right)^{-1}
    \label{eqn:alpha}
\end{equation}
where $n$ is the number of stars and $X=\frac{x_\mathrm{max}}{x_\mathrm{min}}$. 

\citet{khalaj_stellar_2013} also show that the error of $\hat{\alpha}$ can be estimated using the negative of the inverse of the Fisher information which gives
\begin{equation}
    \sigma(\hat{\alpha})=\frac{1}{\sqrt{n}}\left[(\hat{\alpha}-1)^{-2}-\ln^2X\frac{X^{\hat{\alpha}-1}}{(1-X^{\hat{\alpha}-1})^2}\right]^{-1/2}.
\end{equation}

Here it is important to make the distinction between the stellar MF and system MF, which are related yet represent different quantities. The system MF refers to the mass distribution of single stars and primary stars of multiple systems, whereas the stellar MF describes the mass distribution of all stars in the population. In this method, we use the system MF to generate the population of single and primary mass stars for binary systems. We then introduce additional low and medium mass stars as secondary stars from flat distributions rather than the same system MF. This results in the binary corrected stellar MF, which describes the distribution of all individual star masses in the cluster. While the system MF is useful in the creation of stellar populations when the properties of binaries are of interest, the stellar MF is generally the preferred description as it fully measures the complete mass distribution. Thus, we convert the best fitting system PDMF into a stellar PDMF by generating a synthetic cluster with a large number of stars to reduce statistical uncertainty, using the system PDMF and binary fraction and then measure the stellar PDMF using Equation \ref{eqn:alpha}. 

The adopted binary pairing method provides a practical framework that preserves simple analytic forms for both the stellar IMF and the binary mass-ratio distribution. It should be noted that drawing secondary masses from a flat mass-ratio distribution introduces a small excess of systems near $m_\mathrm{min}$ at high binary fractions. However, we find that even for star counts of $N_\mathrm{star}\approx10000$ with 100\% binary fractions, that the small deviation from a perfect power law is dominated by the scatter from stochastic sampling and the resulting stellar mass function retains its overall three-stage power law form with the same break masses. Quantitatively, the introduced deviations are smaller than the stochastic scatter expected from finite sampling, with KS tests indicating that synthetic populations remain consistent with a three-stage power law description. While alternative pairing schemes are possible \citep[e.g.][]{kouwenhovenExploringConsequencesPairing2009, belloniInitialBinaryPopulation2017}, the adopted approach provides a simple, computationally efficient framework that can be described analytically and adequately reproduces the stellar populations modelled in this work. 

\begin{figure}
    \centering
    \includegraphics[width=1\columnwidth]{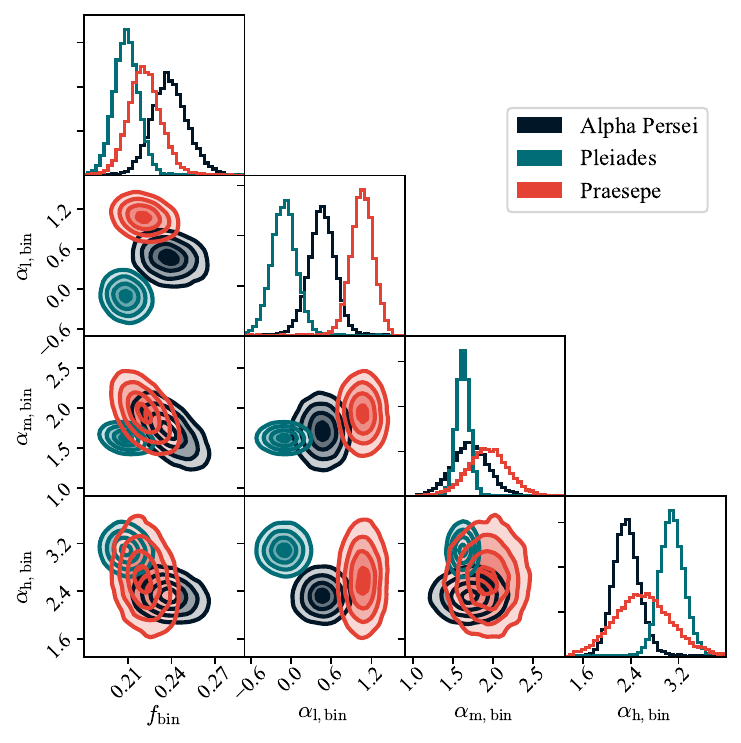}
    \caption{Posterior distributions for the present day binary fraction and three binary corrected system PDMF slopes for Alpha Persei, Pleiades and Praesepe. All three clusters have comparable binary fractions $f_\mathrm{bin}$, high MF slopes $\alpha_\mathrm{h,bin}$ and intermediate MF slopes $\alpha_\mathrm{m,bin}$ but have distinct low MF slopes $\alpha_\mathrm{l,bin}$. }
    \label{fig:bin_cornerplot}
\end{figure}

Figure~\ref{fig:bin_cornerplot} shows the posterior distribution of each cluster. We see that there is good convergence for the estimates of the system MF and binary fraction. The best fitting parameters given by these distributions are listed in Table~\ref{tab:binary_corrected}. We find that the calculated binary fractions are compatible with being the same for all three clusters. Furthermore, the intermediate mass and high mass MF slopes are also compatible, with the low mass slopes significantly deviating from one another.

\begin{table}
    \centering
    \begin{tabular}{cccc}
    Value & Alpha Persei & Pleiades & Praesepe \\
    \hline
    $f_\mathrm{bin}$ & $(23.8\pm1.2)\%$ & $(20.0\pm0.8)\%$ & $(22.1\pm1.0)\%$\\
    $\alpha_\mathrm{h,bin}$ & $2.37\pm0.20$ & $3.11\pm0.19$ & $2.66\pm0.45$\\
    $\alpha_\mathrm{m,bin}$ & $1.76\pm0.21$ & $1.69\pm0.08$ & $2.01\pm0.24$\\
    $\alpha_\mathrm{l,bin}$ & $0.90\pm0.19$ & $0.46\pm0.18$ & $1.41\pm0.16$\\
    \end{tabular}
    \caption{Binary corrected cluster IMF properties. Shown are the estimated binary fraction ($f_\mathrm{bin}$) and the high, intermediate, and low mass PDMF slopes ($\alpha_\mathrm{h,bin}$, $\alpha_\mathrm{m,bin}$, $\alpha_\mathrm{l,bin}$) after correcting for unresolved binaries. Reported uncertainties correspond to 1$\sigma$ errors.}
    \label{tab:binary_corrected}
\end{table}

The binary fractions inferred here for Alpha Persei, the Pleiades, and Praesepe are broadly consistent with values reported in existing studies. For Alpha Persei, we measure $f_\mathrm{bin}=0.238\pm0.012$. Using similar Monte Carlo modelling, \citet{sheikhi_binary_2016} found a binary fraction of $0.23\pm0.09$ across the mass range $0.1-4.7\,\mathrm{M}_\odot$, which is in good agreement with our result. Our estimate for the Pleiades, $f_\mathrm{bin}=0.200\pm0.008$, lies between reported values in previous works. For example, the photometric analysis of \citet{liu_photometric_2025} finds a higher binary fraction of $0.34\pm0.02$, while \citet{cordoni_photometric_2023} report a lower value of $0.148\pm0.100$. Given the different methodologies and sensitivities to binary mass ratios in these studies, our result falls comfortably within the range of existing estimates. Finally, for Praesepe we obtain $f_\mathrm{bin}=0.221\pm0.010$, which is below some prior estimates such as the $0.35\pm0.05$ binary fraction inferred in \citet{khalaj_stellar_2013}, but more consistent with other photometric studies such as the fraction of $0.251\pm0.108$ determined in \citet{cordoni_photometric_2023}. The differences in estimations likely arise due to the various observational methods, stellar mass ranges and detectable mass-ratio limits across different studies. 

\begin{figure}
    \centering
    \includegraphics[width=0.9\columnwidth]{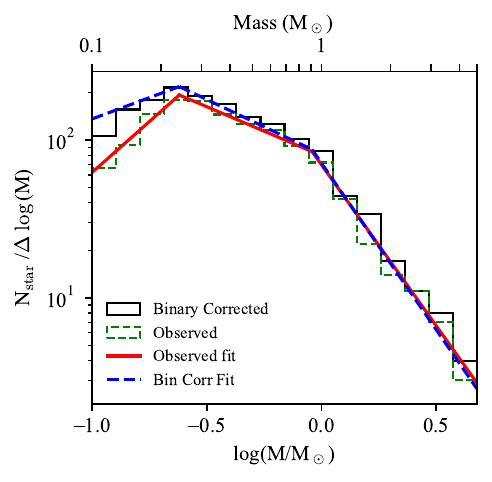}
    \caption{Binary corrections of the PDMF of Pleiades. Binary correction leads to a relatively larger fraction of low and intermediate mass stars compared to high mass stars due to additional, unseen low mass stars that are binary companions.}
    \label{fig:bin_PDMF}
\end{figure}

The effects of binaries on the Pleiades mass function are visually seen in Figure \ref{fig:bin_PDMF}. Factoring in unresolved binaries is shown to flatten the mass function slope in the low mass range and slightly steepen the slope for intermediate mass stars with little effect at the high mass end. This behaviour is expected, and the fact that we see the same effect helps confirm the validity of the results using this Bayesian approach. We see only slight variations in the intermediate mass range, meaning the measurement of the MF for stellar samples between $\gtrsim0.25\,\mathrm{M}_\odot$ are not significantly affected by binary or multiple star contaminants. The largest effect is on the MF below $\approx0.25$, which already has additional uncertainty due to the dependence on the mass estimation model used. 

\section{Cluster initial mass functions}
\label{sec:simulations}

As discussed above, open clusters are very useful laboratories for IMF studies, especially in measuring high mass stellar populations since they are young enough for the most massive stars to remain on the main sequence \citep[e.g.][]{kroupa_variation_2001}. However, their compact size and low masses mean that they have typical relaxation times of the order of 100 Myr. This means that the effects of mass segregation and the preferential stripping of low mass stars on the cluster MF will become apparent after only approximately 100 Myr, comparable to the ages of the clusters we study in this work. Thus, for an accurate determination of the IMF in more evolved open clusters, we must account for the effect of their dynamical evolution on the PDMF.     

To determine the true IMF of each cluster, we used \textit{N}-body simulations to find the initial number of systems, binary fraction, half-mass radius, density profile and mass function that best recreate observations. We vary 7 input parameters that correspond to these properties and compare the final state of different initial parameter combinations to observables of the cluster such as the current binary fraction and PDMF. These parameters were the initial number of systems $N_\text{sys,ini}$, binary fraction $f_\mathrm{bin, ini}$, 3D half-mass radius $r_\mathrm{h,ini}$, a \citet{king_structure_1962} density profile cluster concentration $c$ and IMF slopes $\alpha_\mathrm{l,sys}$, $\alpha_\mathrm{m,sys}$ and $\alpha_\mathrm{h,sys}$. We used the same break points $m_{x1}$ and $m_{x2}$ as the observed clusters rather than leaving them as free parameters to reduce the complexity of the model. This is reasonable since, while the power-law exponents can change due to dynamical evolution, we find the break points remain relatively stable over the cluster's evolution. Even though these simulations are generally quick due to the low number of stars in open clusters, a fixed grid search over 7 parameters is not realistically possible due to the large amount of computation time needed. Even with 10 values per parameter with an optimistic runtime of 5 minutes per parameter combination, a complete search for one cluster would take around a century of computation time. In reality, with initial binary systems implemented, runtimes can vary and be much longer on average, increasing this total time even further. Thus, we implement machine learning and train a model on a coarse grid of \textit{N}-body simulations to significantly reduce computational requirements. 

\subsection{Emulator training}
We ran 942, 672 and 550 complete \textit{N}-body simulations for Alpha Persei, Pleiades and Praesepe respectively using NBODY7 \citep{nitadori_accelerating_2012}. The initial positions and velocities of the simulated clusters were found using the observed central positions, proper motions and radial velocities listed in Table \ref{tab:astro_summ}, and tracing back the path of the cluster in the galactic potential from \cite{irrgang_milky_2013} using a Leapfrog integrator in GALPY \citep{bovy_galpy_2015}. The clusters were generated with a \cite{king_structure_1962} density profile which can be described using a 3D half mass radius $r_\mathrm{h,ini}$ and a dimensionless concentration parameter $c$. For a King model, the tidal radius $r_t$ is related to the core radius $r_\mathrm{c}$ through
\begin{equation}
    r_t=r_\mathrm{c}10^c.
\end{equation}

The secondary masses for primordial binaries were sampled from a flat distribution of mass ratios, similar to the sampling described in Section~\ref{sec:binaries}. Semi-major axes were sampled from a log-uniform distribution bounded between a minimum separation of 50 times the combined stellar radius and up to 1000 AU. These limits would therefore exclude any very close binary systems and wide binary pairs from the initial cluster state in favour of numerical stability and efficiency. Thus, the true initial fraction of binaries may be larger than that explicitly represented within the adopted separation range, since the imposed limits preferentially exclude the closest and widest systems that contribute to the full observationally inferred binary population. Alternative prescriptions for constructing primordial binary populations, including different pairing algorithms and orbital parameter distributions motivated by observations and star formation models, have also been proposed \citep[e.g.][]{kouwenhovenExploringConsequencesPairing2009, moeMindYourPs2017, belloniInitialBinaryPopulation2017}.

We only proceeded with a simulation when the resulting tidal radius was smaller than its Jacobi radius at the cluster's initial galactic radius. For computational efficiency, the Jacobi radius $r_J$ was calculated assuming a circular galactocentric orbit and a flat rotation curve, giving
\begin{equation}
    \label{eqn:jacobi}
    r_J = \left(\frac{GM_cR_g^2}{2v_c^2}\right)^{1/3},
\end{equation}
where $M_c$ is the cluster mass, $R_g$ is the galactic centred radius and $v_c$ is the circular velocity of the Galaxy at that radius.

This condition ensures that the cluster initially lies within its tidal boundary in the Galactic potential, preventing stars at large radii from being immediately stripped by the galactic tidal field. Without this, the input parameters would not represent a physically stable initial cluster configuration, affecting the relationships between input and output parameters that the machine learning models are trained on. We also assume that all neutron stars and black holes are ejected from the cluster due to the large velocity kicks at their creation and the relatively low escape velocity of the investigated clusters \citep{pavlik_black_2018}. 

Initial conditions were sampled from a 7-dimensional Latin hypercube spanning a broad range chosen to encompass all physically reasonable initial conditions, ensuring adequate coverage of parameter space for emulator training. If the best fitting parameters were later found to be in a region with a low density of training data, we ran additional simulations in that region of parameter space. Once all simulations were completed, we applied the same $\chi^2$ filtering described in Equations \ref{eq:chi2} and \ref{eqn:chi_cond} on the final state of the simulated clusters so that the present day values in the simulations could be compared with observations. 

We then measured the final number of systems, binary fraction and 2D half mass radius of each simulated cluster as well as the mass function using the same break points as the observed clusters. The surface density of each cluster was also measured by binning stars into 10 radial bins with an equal number of stars in each, and finding 
\begin{equation}
    \rho_\mathrm{surf,i} = \frac{N_\mathrm{star,i}}{\pi(r_i^2-r_{i-1}^2)}
\end{equation}
for each bin $i$ where $r_i$ is the outer radius of bin $i$. 

{
\footnotesize
\begin{table*}
    \centering
    \begin{tabular}{lccccccccc}
    Cluster & $N_\text{sys,ini}$ & $f_\mathrm{bin, ini}$ & $r_\mathrm{h,ini}$ (pc) & $\alpha_\mathrm{l,sys}$ & $\alpha_\mathrm{m,sys}$ & $\alpha_\mathrm{h,sys}$ & $\alpha_\mathrm{l,star}$ & $\alpha_\mathrm{m,star}$ & $\alpha_\mathrm{h,star}$ \\
    \hline
    Alpha Persei & $1127^{+30}_{-49}$ & $0.346^{+0.019}_{-0.014}$ & $5.63^{+0.39}_{-0.55}$ & $0.48\pm0.23$ & $1.58\pm0.25$ & $2.60\pm0.19$ & $0.98\pm0.23$ & $1.64\pm0.25$ & $2.60\pm0.19$\\
    Pleiades & $1670^{+44}_{-28}$ & $0.303\pm0.012$ & $4.45^{+0.22}_{-0.33}$ & $0.09\pm0.27$ & $1.58\pm0.09$ & $3.36\pm0.10$ & $0.80\pm0.27$ & $1.67\pm0.09$ & $3.33\pm0.10$\\
    Praesepe & $2105^{+187}_{-79}$ & $0.242\pm0.015$ & $4.34^{+0.49}_{-0.95}$ & $1.57\pm0.24$ & $2.15\pm0.26$ & $2.86^{+0.27}_{-0.36}$ & $1.93\pm0.24$ & $2.23\pm0.26$ & $2.86\pm0.36$\\
    \end{tabular}
    \caption{Best fitting initial cluster parameters. Reported values include the inferred initial number of stellar systems ($N_\text{sys,ini}$), initial binary fraction ($f_\mathrm{bin,ini}$), initial 3D half-mass radius ($r_\mathrm{h,ini}$), the system IMF slopes for the low, intermediate and high mass ranges ($\alpha_\mathrm{l,sys}$, $\alpha_\mathrm{m,sys}$, $\alpha_\mathrm{h,sys}$) and the stellar IMF slopes for the low, intermediate and high mass ranges ($\alpha_\mathrm{l,star}$, $\alpha_\mathrm{m,star}$, $\alpha_\mathrm{h,star}$). Uncertainties correspond to 1$\sigma$ errors.}
    \label{tab:dynamic_corrected}
\end{table*}
}

We then found the surface density of each simulation's final state using the same radial bins as the observed clusters. Thus, in total we have 16 test statistics that we can use to compare the simulated clusters to the observations. The aim of this analysis is to create an emulator that will take any combination of the 7 simulation input parameters and give a prediction of these 16 test statistics without needing to process an entire \textit{N}-body simulation of the cluster. 

To create this emulator and find the optimal method of emulation, we use the python package AUTOEMULATE \citep{stoffel_autoemulate_2025}. Using this tool, we compare various models to find the method which maximises the $R^2$ statistic when applied to a test subset of data. From this analysis, we found that using a correlated Gaussian process was optimal for modelling the relationships between the input parameters and test statistics. 

On initial inspection, we found that the concentration parameter $c$ had little significant correlation with any observable test statistic and was best tuned so that the initial cluster had a tidal radius approximately equal to the Jacobi radius for a given initial half mass radius. Thus, we omitted the input variable $c$ in the emulator training model to reduce complexity in the emulator and the subsequent solver. 

\subsection{Parameter fitting}

To find the initial cluster conditions that are most likely to recreate observations, we used the inbuilt Bayesian modelling in AUTOEMULATE. More specifically, we assume that all observables have Gaussian errors and implement a Hamiltonian Monte Carlo (HMC) No-U-turn sampler (NUTS) algorithm to sample the posterior distribution. We then report the median values of the sample as the best fitting values with uncertainties given by the 16th and 84th percentiles. The resulting posterior distributions can be seen in Figure \ref{fig:comb_corner} and the best fitting values in Table \ref{tab:dynamic_corrected}. 

\begin{figure*}
    \centering
    \includegraphics{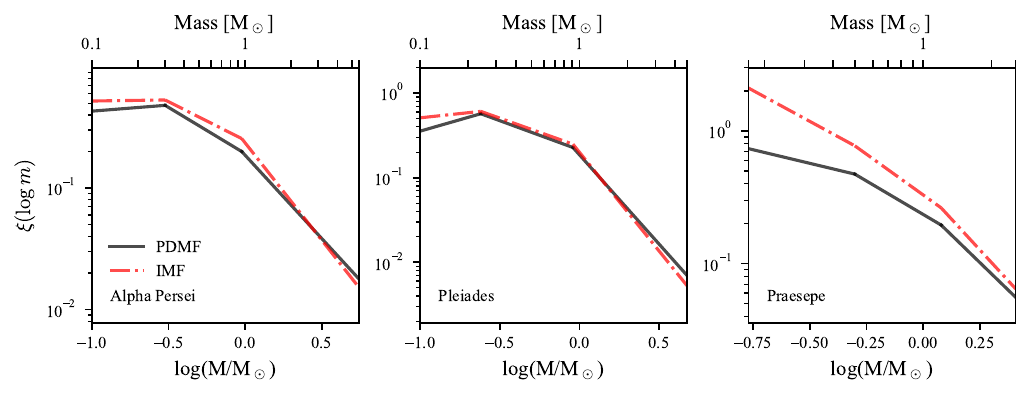}
    \caption{The derived highest likelihood stellar IMF (red) for each cluster compared to its PDMF (black). The mass function $\xi(\log m)$ is scaled such that the curves reflect simulated stellar losses of $\approx15\%$, $9\%$ and $49\%$ for Alpha Persei, Pleiades and Praesepe respectively. We see that Praesepe is more evolved and thus has a larger number of low mass stars that have been stripped over time, whereas Alpha Persei and Pleiades are younger and have not gone through as much significant dynamical evolution.}
    \label{fig:IMF_slopes}
\end{figure*}

Due to the way the cluster is created for the simulations, we solved for the maximum likelihood estimators of the initial system IMF and binary fraction rather than the initial stellar IMF. Similarly to Section \ref{sec:synth_clusters}, we create a large synthetic cluster to estimate the corresponding stellar IMF of each cluster. Using this method, we find the true IMF of the three clusters in Table \ref{tab:dynamic_corrected} and Figure \ref{fig:IMF_slopes}. 

We see that Alpha Persei and Pleiades undergo minimal dynamical evolution which is consistent with expectations due to their relatively low age with respect to their relaxation times. This is in contrast to Praesepe, which is seen to have lost a significant fraction of lower mass stars below $\approx0.6\,\mathrm{M}_\odot$ due to mass segregation and prolonged tidal stripping. Interestingly, we find that the high mass IMF of each cluster is steeper than the observed PDMF by about 0.2. This is likely driven more by the dependence of the final cluster state's half mass radius on the initial high mass function slope as a larger number of high mass stars will cause additional dynamical heating as well as drive stronger expansion early in the cluster's evolution from the shallower gravitational potential caused by early, high mass star evolution. Thus, the slope of the IMF at high masses is constrained on both the measurement of the high mass PDMF and the present day half mass radius.

\subsection{Validation of initial conditions}
We verified the best fitting parameters of each cluster using \textit{N}-body simulations and comparing the final states to the observed clusters. Since every created cluster and simulation will have random variations, especially with the smaller sample of stars in open clusters, we ran at least 20 simulations with the best fitting parameters for each cluster and chose the run with the best agreement with observations. 

\begin{figure*}
    \centering
    \includegraphics{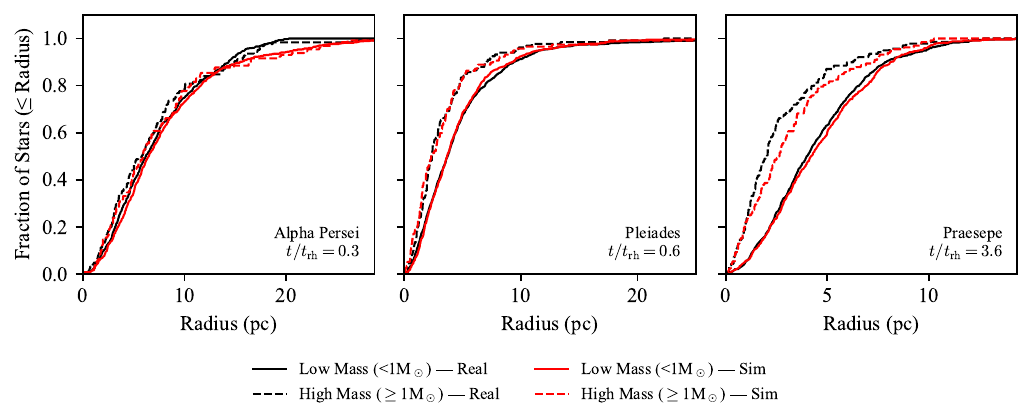}
    \caption{Cumulative distribution of stars in each cluster with respect to radius (black). Cluster members are split into two groups, showing the different radial distribution of high mass stars $>1\text{M}_\odot$ and low mass stars $<1\text{M}_\odot$ which indicates the presence of mass segregation. We also show the radial distribution of simulations using the best fitting parameters (red) which are in good agreement, despite mass segregation not being used in the goodness of fit estimation.}
    \label{fig:mass_seg}
\end{figure*}

We find that the simulations are able to reproduce the observed number of stars, binary fraction, half mass radius and three mass function slopes within $2.5\sigma$, with most parameters agreeing within $1\sigma$. In addition, while we do not include any indicators of mass segregation into the optimised likelihood function, we find good agreement in the degree of mass segregation between observations and simulations. A  Kolmogorov-Smirnov (KS) test between each of the observed and simulated radial distributions found no statistically significant differences, with p-values ranging between 0.15 – 0.77. As shown in Figure \ref{fig:mass_seg}, the degree of mass segregation for each cluster is related to their age with respect to their relaxation times. The difference in the observed low mass and high mass radial distributions in Alpha Persei shows no statistically significant difference, with $p=0.30$, whereas Pleiades and Praesepe show significant differences in their stellar radial distributions with $p<10^{-5}$ and $p<10^{-15}$ respectively.

\begin{figure}
    \centering
    \includegraphics[width=0.9\columnwidth]{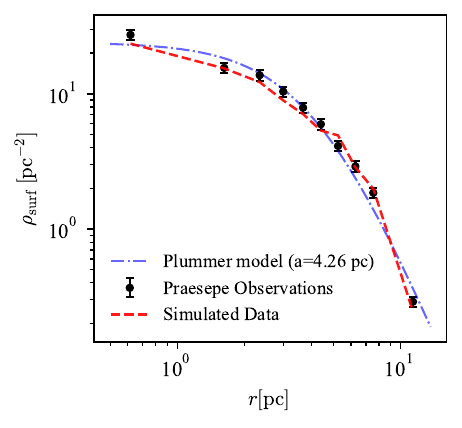}
    \caption{The observed surface density profile of Praesepe (black) and the surface density profile of a simulation using the best fitting parameters (red dashed). There is good agreement between the two curves, showing the validity of the derived parameters. We also show the best fitting Plummer model for the observed surface density (blue dot-dashed) with Plummer radius $a=4.26\pm0.11$ and $N_\mathrm{star}=1367\pm40$. }
    \label{fig:sim_surf_dens}
\end{figure}

We also find good agreement with the simulated and observed surface density profiles for each cluster, with Praesepe shown in Figure \ref{fig:sim_surf_dens}. \citet{roser_praesepe_2019} found that the surface density of Praesepe is well fitted with a Plummer model \citep{plummer_problem_1911} with core radius $r_\mathrm{co}=3.7\,\mathrm{pc}$. We fit a 2D Plummer model to the surface number density of each cluster, given by
\begin{equation}
    \label{eqn:plummer}
    \rho_\mathrm{surf}(r) = \frac{N_\mathrm{star}}{\pi a^2}\left(1+\frac{r^2}{a^2}\right)^{-2},
\end{equation}
where $N_\mathrm{star}$ is the total number of stars in the cluster and $a$ is the Plummer radius. 

We fit both $N_\mathrm{star}$ and $a$ as free parameters for each cluster using a non-linear least squares fit. For Alpha Persei, we find the best fitting Plummer model is given by $a=7.17\pm0.20$ pc and $N_\mathrm{star}=1105\pm35$ which has a reduced $\chi^2$ value of $\chi^2_\mathrm{red}=2.35$. The surface density profile of Pleiades is fitted by $a=4.13\pm0.10$ pc and $N_\mathrm{star}=1646\pm42$ with $\chi^2_\mathrm{red}=8.48$ and Praesepe is described best by a model with $a=4.26\pm0.11$ pc and $N_\mathrm{star}=1367\pm40$ with $\chi^2_\mathrm{red}=3.57$. While all three models give $\chi^2_\mathrm{red}>1$, the Plummer model still provides a reasonable approximation of the surface density profile of each cluster using a simple analytical model. This can be useful for obtaining a first-order estimate of the structural scale of each cluster and for providing a simple analytic description of the observed stellar distributions. 

The agreement between these forward simulation runs and cluster observations help demonstrate that these solutions are valid initial conditions of the three investigated clusters. These values, however, are not necessarily the only possible initial state of the cluster. \citet{wang_impact_2021} demonstrate that small changes in the number of high mass OB stars have significant effects on the evolution of a cluster over the span of a few 100 Myrs. If we assume that the IMF is randomly sampled during star formation and acts as a probability distribution function, similar to the sampling methods of this work, then stochastic variations in the number of high-mass stars can significantly alter the subsequent dynamical evolution and present-day state of a cluster compared to one drawn from the same IMF but containing fewer massive members. There are also additional variations introduced due to random dynamical encounters as well as the random initialisation of binaries in each \textit{N}-body simulation. We expect all of these stochastic features to introduce scatter in the training data which the emulator models are trained on, however, the degree of variations should be reflected by uncertainties of the emulators predicted values. These uncertainties will then be reflected by a wider spread of data when sampling the posterior distribution and thus a larger uncertainty in the predicted initial parameters. 

Thus, in this work we have found gas-free initial conditions of the present-day cluster state that have the highest likelihood of reproducing present-day observations, but will have variations with each realisation of the cluster. While the true initial conditions may vary, we show that this method is able to accurately recreate observations and provides a useful initial state for simulation work that wishes to study the evolution of a cluster under different conditions.

\subsection{Comparison to other works}

\begin{figure}
    \centering
    \includegraphics[width=\columnwidth]{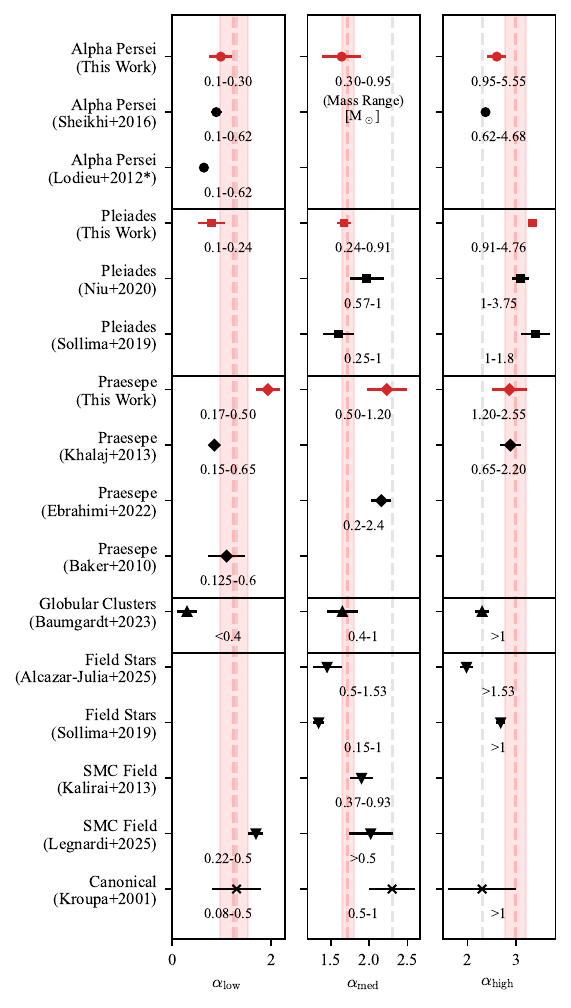}
    \caption{Comparison of IMF measurements of Alpha Persei, Pleiades, Praesepe and other stellar bodies. The mass range in M$_\odot$ which the value was calculated over is noted underneath each data point. The red vertical line and shaded region depict the mean IMF slope values for our three measured clusters with uncertainty. The grey vertical line corresponds to the slope values from the canonical \citet{kroupa_variation_2001} IMF.}
    \label{fig:lit_comp}
\end{figure}

While a direct comparison between different IMF measurements is difficult due to the different mass ranges which the mass function slopes are calculated over, we compare our results to other IMF determinations of the same three clusters as well as other system types for similar mass ranges. As shown in Figure \ref{fig:lit_comp}, our IMF values agree with most other determinations of cluster IMFs over similar mass ranges with only a few exceptions. The largest discrepancies are likely due to the significantly different mass ranges which the power law is calculated over. In many studies of star cluster IMFs, the adopted break points for the multi-stage power law are ones that give a global best fit, rather than being specific to each cluster. While this fits into the framework of a universal cluster IMF, it can result in a IMF that does not well describe all of the features in a particular cluster and can smooth out unique turnover behaviour, resulting in what appears to be a universal average mass function slope that is not necessarily a very good fit for a number of clusters. 

In our approach, we determine the break points for each cluster independently and find that a high mass function break of $m_{x2}=1\,\mathrm{M}_\odot$ is within error bars for all three clusters, consistent with the majority of other IMF determination studies. The low-intermediate mass break point $m_{x1}$, however, is less defined and has greater variation between the three clusters as well as with other studies of the same clusters. While there could be a real variation of turn over behaviour at low masses between clusters, the specific values of $m_{x1}$ we find are highly dominated by the choice of mass-luminosity relation or isochrone, meaning we cannot confidently determine any physical relationships. 

We find nearly all our individual cluster measurements to be in general agreement with other determinations over similar mass ranges. For Alpha Persei, \citet{sheikhi_binary_2016} find that $\alpha_\mathrm{l}=0.89\pm0.11$ and $\alpha_\mathrm{h}=2.37\pm0.09$ using the PPMXL and \textit{Wide-field Infrared Survey Explorer} catalogues, correcting for unresolved binaries, and \citet{lodieu_astrometric_2012} find that $\alpha_\mathrm{l}=0.64\pm0.08$ using UKIDSS data. These values are in good agreement with our findings of $\alpha_\mathrm{l}=0.98\pm0.23$, $\alpha_\mathrm{m}=1.64\pm0.25$ and $\alpha_\mathrm{h}=2.60\pm0.19$. The slight difference in the high-mass function slope likely arises from our use of two power laws over 0.30–0.95 M$_\odot$ and 0.95–5.55 M$_\odot$, rather than a single power law over 0.62–4.68 M$_\odot$ as adopted by \citet{sheikhi_binary_2016}. Our IMF slopes for Pleiades in the intermediate and high mass ranges also agree with works such as \citet{niu_binary_2020} and \citet{sollima_stellar_2019} who find $\alpha_\mathrm{m}=1.97\pm0.22,1.6\pm0.2$ and $\alpha_\mathrm{h}=3.09\pm0.18,3.4\pm0.3$ respectively using LAMOST and \textit{Gaia} DR2 observations. 

Lastly, we also find good agreement with previous determinations for Praesepe for the high and intermediate MF slopes. \citet{khalaj_stellar_2013} found that $\alpha_\mathrm{h}=2.88\pm0.22$ which agrees well with our measured value of $\alpha_\mathrm{h}=2.86\pm0.36$. Our measured intermediate IMF slope of $\alpha_\mathrm{m}=2.23\pm0.24$ also agrees well with the findings of \citet{ebrahimi_family_2022} who determined $\alpha_\mathrm{m}=2.16\pm0.13$, although these values are all calculated over slightly different mass ranges which may lead to slightly different values. For the low mass IMF slope, however, we find a much steeper value compared to published values. \citet{khalaj_stellar_2013} found that $\alpha_\mathrm{l}=0.85\pm0.10$, and \citet{baker_low-mass_2010} determined that $\alpha_\mathrm{l}=1.10\pm0.37$ using near infrared survey data, whereas we find a higher value of $\alpha_\mathrm{l}=1.93\pm0.24$. The main contributing factor to this discrepancy is likely our inclusion of dynamical evolution which would include many additional low mass stars compared to the present-day state of the cluster. Moreover, there is additional error in the exact slope value due to the uncertainties of the mass conversion models for these low masses as previously discussed.

In addition, we also find that simulations of Praesepe which use the best fitting parameters can qualitatively reproduce the tidal tails observed by \citet{roser_praesepe_2019}. In our simulations, we identify a total of 700 stars in the tidal tails outside of two tidal radii and within a radius of 165 pc, matching the observational bounds of \citet{roser_praesepe_2019} who used \textit{Gaia} DR2 data. Of these stars, we find 339 in the leading tail and 361 in the trailing tail. This is broadly consistent with the 389 tail stars identified by \citet{roser_praesepe_2019}, particularly considering that observations of the trailing arm are likely incomplete for fainter stars due to its larger distance from the Sun. The overall agreement suggests that our simulations are able to capture the dominant dynamical processes responsible for the formation of these extended tidal features. 

We note, however, that our simulations assume gas-free initial conditions. As shown by \citet{weis_dynamical_2025}, the inclusion of instantaneous gas expulsion at cluster birth can remove up to $\approx75\%$ of the initial cluster mass, implying a significantly larger initial stellar population than considered here. In the work of \citet{weis_dynamical_2025}, this leads to a considerably different inferred IMF of Praesepe, with a very steep high mass slope ($\alpha_\mathrm{h}=4.1$) and much fewer low mass stars ($\alpha_\mathrm{l}=0.4$). However, their analysis does not consider unresolved binaries which, as discussed in Section~\ref{sec:binaries}, can bias the inferred mass function. As such, the differences between these findings reflect both differing initial assumptions and modelling choices. We also note that a range of similar dynamical modelling approaches have previously been applied to star clusters. For example, \citet{kroupa_formation_2001} modelled the early gas-embedded evolution of a Pleiades-like cluster from Orion Nebula Cluster-like initial conditions, while \citet{haghi_possible_2015} show the effect of gas expulsion on the MF in the first 100 Myr of the evolution of globular clusters. However, \citet{haghi_possible_2015} found that gas expulsion alone does not strongly alter the inferred IMF unless strong primordial mass segregation is also present, which we do not include in our models.

To compare our results to the mass functions of other stellar groups, we found the mean IMF slope of the three open clusters for each mass range. We also determined the standard deviation of each IMF slope similar to the formalism in \citet{dib_massive_2017}, where the IMF has cluster-to-cluster variations that can be described by Gaussian distributions. From the three clusters investigated, we find the average IMF to be given by $\bar\alpha_l=1.24\pm0.29$, $\bar\alpha_m=1.72\pm0.09$ and $\bar\alpha_\mathrm{h}=2.98\pm0.22$ with dispersions of $\sigma_\mathrm{l}=0.43\pm0.24$, $\sigma_\mathrm{m}=0.00\pm0.14$, $\sigma_\mathrm{h}=0.29\pm0.16$ and break masses of $m_{x1}=0.25\pm0.02\,\mathrm{M}_\odot$ and $m_{x2}=0.96\pm0.14\,\mathrm{M}_\odot$. 

\begin{figure}
    \centering
    \includegraphics[width=0.9\columnwidth]{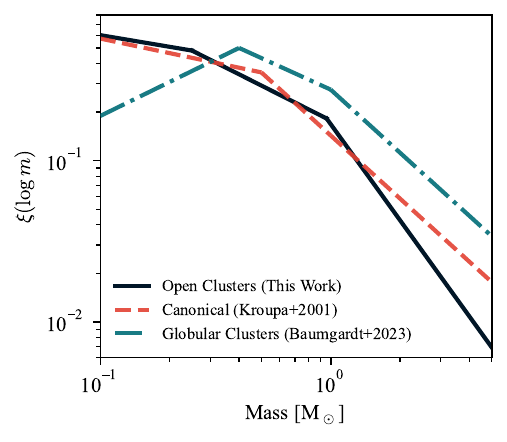}
    \caption{Average open cluster initial mass function (IMF) (black solid line) compared to the canonical IMF from \citet{kroupa_variation_2001} (red dashed) and the average globular cluster IMF from \citet{baumgardt_evidence_2023} (blue dot-dashed). All three curves are normalised such that the area under the curve is unity. The open cluster IMF closely follows the canonical shape but exhibits a flattening at intermediate masses and a steepening at high masses, implying a relative deficiency of massive stars. Relative to the globular cluster IMF, it contains a larger fraction of low-mass stars and fewer high-mass stars. }
    \label{fig:litcompplots}
\end{figure}

This average open cluster IMF is shown in Figure~\ref{fig:litcompplots} against the canonical field IMF from \citet{kroupa_variation_2001} and the measured average IMF of globular clusters as determined in \citet{baumgardt_evidence_2023}, where all functions are normalised so that the area under the IMF is unity. In comparison to both functions, the open cluster IMF depicts a smaller population of high mass stars $>1\,\mathrm{M}_\odot$. While the values of the break points and slopes are different to those quoted in \citet{kroupa_variation_2001}, the turnover shape of the IMF for low and intermediate mass stars below $\approx1\,\mathrm{M}_\odot$ is qualitatively similar. This is in contrast to the bottom-light IMF form observed in globular clusters which have a significantly smaller proportion of low mass stars $<0.3\,\mathrm{M}_\odot$ and larger proportion of intermediate and high mass stars. The apparent deficit high mass stars in the investigated metal-rich open clusters compared to metal-poor, denser globular clusters is in qualitative agreement with previous studies that suggest denser star forming environments produce increasingly top-heavy IMFs \citep{marks_evidence_2012, dabringhausenLowmassXRayBinaries2012}.

\section{Discussion}

Our results indicate that the open clusters studied have an IMF that is top light compared to IMF measurements of field stars and star clusters. The average high mass IMF slope of $\bar\alpha_\mathrm{h}=2.98\pm0.22$ for the three open clusters studied sits on the very upper end of uncertainty estimates of field star estimates from \citet{sollima_stellar_2019} who find an average slope above $1\,\mathrm{M}_\odot$ of $\alpha_\mathrm{h}=2.68\pm0.09$ as well as the canonical slope measurement of $\alpha_\mathrm{h}=2.3\pm0.7$ from \citet{kroupa_variation_2001}. We do, however, find much stronger discrepancies between this value with other determinations of the IMF over similar mass ranges in field stars and globular clusters, with \citet{alcazar-julia_joint_2025} finding $\alpha_\mathrm{h}=1.98^{+0.13}_{-0.05}$ using \textit{Gaia} measurements of stars in the Milky Way disk as well as \citet{baumgardt_evidence_2023}, who found the IMF slope over $1\,\mathrm{M}_\odot$ of LMC and SMC clusters to be $\alpha_\mathrm{h}=2.30\pm0.15$. The comparatively steeper slope we infer therefore suggests that, if robust against remaining systematic uncertainties, star formation in low-mass open clusters may be intrinsically deficient in the highest-mass stars relative to the broader galactic field and massive cluster populations.

For the intermediate mass slope, we find a mean value of $\bar\alpha_\mathrm{m}=1.72\pm0.09$ which appears much shallower than the traditionally used canonical value of $\alpha_\mathrm{m}=2.3\pm0.3$ \citep{kroupa_variation_2001} but is more in line with recent measurements of field stars and globular clusters. \citet{alcazar-julia_joint_2025} and \citet{sollima_stellar_2019} determined MF slope values over similar mass ranges of $\alpha_\mathrm{m}=1.45^{+0.19}_{-0.12}$ and $\alpha_\mathrm{m}=1.34\pm0.07$ respectively. Furthermore, SMC field star measurements from \citet{kalirai_ultra-deep_2013} and \citet{legnardi_small_2025}, which use deep infrared measurements from the HST and JWST, find similar values of $\alpha_\mathrm{m}=1.90^{+0.15}_{-0.10}$ and $\alpha_\mathrm{m}=2.02\pm0.29$ respectively. Lastly, \citet{baumgardt_evidence_2023} find an average intermediate IMF slope value of $\alpha_\mathrm{m}=1.65\pm0.20$ for globular clusters. The agreement between these independent measurements across field stars, open clusters, and globular clusters therefore suggests that an intermediate-mass slope near $\alpha_m \approx 1.5$–$1.8$ may more accurately represent the typical stellar IMF in this mass regime. Furthermore, this broad consistency across environments indicates that the intermediate-mass regime may be universal and that star formation in this mass region could be relatively insensitive to environmental conditions. 

We find an average slope for low mass stars of $\bar\alpha_\mathrm{l}=1.24\pm0.29$ which appears to be consistent with the canonical value of $\alpha_\mathrm{l}=1.3\pm0.5$, although this conclusion is limited by both the spread in measured values between the different clusters as well as the inaccuracy introduced into the model from mass-luminosity conversion models. As discussed in Section~\ref{sec:PDMF}, the measurement of the PDMF in low mass regime is strongly dependent on the choice of the model when converting the photometric observations to mass estimates. Consequently, more robust conclusions about the universality or variation of the cluster IMF at low masses require improved theoretical mass-luminosity relations and deeper observational constraints. The future release of \textit{Gaia} DR4 is expected to significantly improve the precision of stellar measurements and refine the relationship between the stellar mass of astrometric binaries and their \textit{Gaia} photometric properties \citep{chevalier_binary_2023}. With the precise measurements of more binary systems using \textit{Gaia}, combined with radial velocity measurements, one would be able to calibrate a much stronger relationship between mass, luminosity and colour which would become an incredibly powerful tool in the study of mass functions using \textit{Gaia}.

Although not definitive, these results provide additional evidence that the IMF of star clusters is not strictly universal and likely depends on environmental conditions. Within uncertainties, the IMF appears broadly consistent with a universal form in the intermediate regime, but deviations at high mass may hint at environmental dependencies of high mass star formation. Theoretical models suggest that the formation of massive stars is favoured in dense cluster environments where high accretion rates can be maintained, whereas the lower densities typical of many open cluster progenitors may limit this growth \citep{bonnell_massive_2004, bonnell_star_2006}. In addition, theoretical studies of molecular clouds further indicate that massive star formation is closely associated with regions of high gas surface density \citep{krumholz_minimum_2008}. If open clusters are indeed systematically characterised by a top-light IMF, this would not only provide insight into the physical conditions governing star formation but would also have broader implications for the interpretation of unresolved stellar populations. Massive stars dominate the luminosity of young stellar systems and strongly influence measurements of star formation histories, meaning that environmentally dependent IMFs could introduce systematic biases into population synthesis models that assume a universal IMF \citep[e.g.][]{conroy_counting_2012, jerabkovaImpactMetallicityStar2018, gunawardhanaGalaxyMassAssembly2011}. 

One possibility, however, is that the current sample of open clusters is affected by a form of survivor bias, where clusters with more Salpeter-like IMF slopes at high masses are more likely to dissolve over shorter timescales due to the previously discussed faster expansion from additional high mass stars. This would make it more likely to observe open clusters with steeper IMF slopes as those clusters are more gravitationally stable and dissolve over much longer timescales. Thus, further detailed measurements of open cluster IMFs are required to confirm whether this behaviour in the IMF at high masses is a result of different environmental conditions or stochastic scatter from random sampling. 

In this paper, we have assumed that star formation is a stochastic process and thus the IMF is sampled randomly like a probability distribution function. In this case, we find that the scatter in open cluster IMF measurements due to stochastic sampling would make any weak to moderate relationships with environmental conditions between clusters difficult to resolve over noise without a large sample of cluster measurements. Even if there was an IMF dependence on environmental conditions such as metallicity or density, the intrinsic scatter of open cluster IMFs would make any precise fundamental relationship challenging to observe and quantify with high accuracy. Furthermore, the dynamical evolution of open clusters is easily perturbed by small random encounters and can result in a very different final mass distribution. As discussed previously, this means that the IMF of open clusters cannot be definitively determined, rather the most probable solution can be found. Thus, any environmental relationships would be affected by additional noise, making measurements less precise, if possible at all.

\section{Summary and Conclusion}

To summarise, in this work we have determined the most probable initial binary fraction, mass function and initial mass of the open clusters Alpha Persei, Pleiades and Praesepe. Using \textit{Gaia} DR3 proper motions and parallaxes, supplemented with UKIDSS and HIPPARCOS observations, we determine membership probabilities for stars down to masses of 0.1-0.17 M$_\odot$ in the different clusters. We use \textit{Gaia} photometry in a Bayesian framework to determine the current binary fractions of each cluster and quantify the effect of binaries on the present-day mass function. Lastly, we ran a grid of \textit{N}-body simulations to train an emulator model for each cluster in order to predict the best fitting initial conditions and mass functions using a Hamiltonian Monte Carlo algorithm. 

\begin{itemize}
    \item We determined the fraction of unresolved binaries in each cluster's stellar population to be $0.238\pm0.012$, $0.200\pm0.008$ and $0.221\pm0.010$ for Alpha Persei, Pleiades and Praesepe respectively.

    \item The initial values for $N_\text{sys,ini}$, $f_\mathrm{bin, ini}$ and $r_\mathrm{h,ini}$ were found simultaneously with the IMF values for each cluster. These show that the initial gas free states of these clusters had between $\approx1100-2100$ systems with around 24-35\% of them being components of a binary system. We also find the clusters to be quite large at the time of their creation, having 3D half mass radii between $\approx4-6$ pc. 

    \item We find the IMF of the three open clusters to be adequately described by a three-stage power law with break points at 0.24-0.50 M$_\odot$ and 0.91-1.20 M$_\odot$. The average IMF slopes for the clusters were found to be $\bar\alpha_l=1.24\pm0.29$, $\bar\alpha_m=1.72\pm0.09$ and $\bar\alpha_\mathrm{h}=2.98\pm0.22$ with some scatter between the individual clusters. We note, however, that the inferred low-mass slope is sensitive to the adopted mass–luminosity relation used to convert photometry into stellar masses, and should therefore be interpreted with caution. This mass function is top-light compared to a Salpeter-like IMF slope of $\alpha_\mathrm{h}=2.35$, meaning they have a relative deficiency in high mass stars above $1\,\mathrm{M}_\odot$. 

    \item The best fitting initial parameters were able to reproduce simulated clusters that agree well with observed cluster parameters within 2.5$\sigma$, with most measured parameters agreeing within $1\sigma$. There was also good agreement with the amount of present-day mass segregation between the simulated and observed clusters, with the degree of segregation clearly increasing with the ratio of cluster age over relaxation time $t_\mathrm{rh}$.
    
\end{itemize}

Our findings provide evidence for a cluster IMF with which is top-light compared to the local field star and canonical IMF and additionally shows some cluster-to-cluster scatter. Any exact dependencies on environmental conditions, however, cannot be accurately determined or quantified from the current sample of open clusters from this work due to the low number of clusters and the significant probabilistic scatter through stochastic star formation and dynamical interactions. Expanding the sample of well-characterised clusters across a wider range of masses, ages, and environments will therefore be essential to determine whether the observed trend reflects a genuine environmental dependence of the IMF. Such an expansion is now becoming feasible with recent \textit{Gaia} based cluster catalogues, which have identified large and homogeneous samples of open clusters \citep[e.g.][]{cantat-gaudin_painting_2020, hunt_improving_2024}. Furthermore, we highlight that improved constraints on mass-luminosity relations at low stellar masses will be essential for reducing systematic uncertainties in the inferred low-mass IMF slope.  Together, these improvements will be crucial for determining whether the IMF in open clusters exhibits genuine environmental variation or remains broadly universal. 

\begin{acknowledgement}

We thank the anonymous referee for suggestions that improved the quality of the manuscript. We also thank Cullan Howlett for helpful discussions and guidance on Bayesian statistics and machine learning methods used in this work. This work has made use of data from the European Space Agency (ESA) mission {\it Gaia} (\url{https://www.cosmos.esa.int/gaia}), processed by the {\it Gaia} Data Processing and Analysis Consortium (DPAC, \url{https://www.cosmos.esa.int/web/gaia/dpac/consortium}). Funding for the DPAC has been provided by national institutions, in particular the institutions participating in the {\it Gaia} Multilateral Agreement. This work was supported by computational resources provided by The University of Queensland via the Friday supercomputer.

\end{acknowledgement}

\section*{Data Availability}

The observational data used in this study, including the member catalogues for the three open clusters analysed, with absolute magnitude and mass estimates are publicly available at: https://lachlanhobart.github.io/Data/

\bibliography{references}

\clearpage
\appendix

\onecolumn
\section{Empirical M-L Relation Data}

\begin{longtable}{lrrrrr}
\caption{Stellar data used to construct the empirical mass–luminosity relation. Columns list the star name, mass (M$_\odot$) with uncertainty, absolute $V$-band magnitude (M$_V$) with uncertainty, and reference identifier for the source.} \\
\label{tab:EncBinaryData} \\

\toprule
Name & Mass (M$_\odot$) & $\Delta$Mass & M$_V$ (dex) & $\Delta \mathrm{M}_V$ & ref \\
\midrule
\endfirsthead

\midrule
Name & Mass (M$_\odot$) & $\Delta$Mass & M$_V$ (dex) & $\Delta \mathrm{M}_V$ & ref \\
\midrule
\endhead

\midrule
\endfoot

\bottomrule
\endlastfoot

AD Boo          & 1.414 & 0.009 & 3.12 & 0.10 & 1 \\
AI Phe          & 1.234 & 0.004 & 3.29 & 0.17 & 1 \\
BD+25 2800      & 1.209 & 0.006 & 4.06 & 0.11 & 1 \\
BD+34 4217      & 0.814 & 0.013 & 6.45 & 0.11 & 1 \\
BD+37 1868      & 1.288 & 0.006 & 3.59 & 0.12 & 1 \\
BD+37 2641      & 0.968 & 0.012 & 4.57 & 0.16 & 1 \\
BD+47 1350      & 0.877 & 0.017 & 5.71 & 0.12 & 1 \\
BD+52 3383a     & 0.907 & 0.017 & 6.04 & 0.15 & 1 \\
BD+62 1132      & 0.969 & 0.005 & 5.02 & 0.12 & 1 \\
BD+80 112       & 1.496 & 0.016 & 3.05 & 0.11 & 1 \\
BD-04 1937      & 1.462 & 0.010 & 3.12 & 0.09 & 1 \\
BH Vir          & 1.166 & 0.008 & 4.05 & 0.11 & 1 \\
CD Tau          & 1.442 & 0.016 & 3.17 & 0.07 & 1 \\
CG Cyg          & 0.941 & 0.014 & 5.59 & 0.22 & 1 \\
CV Boo          & 1.032 & 0.013 & 4.32 & 0.15 & 1 \\
EW Ori          & 1.174 & 0.012 & 4.37 & 0.10 & 1 \\
FL Lyr          & 1.218 & 0.016 & 3.95 & 0.11 & 1 \\
GSC 04619-00585 & 1.081 & 0.007 & 4.03 & 0.11 & 1 \\
HD 121909       & 1.052 & 0.006 & 4.81 & 0.22 & 1 \\
HD 124784       & 0.854 & 0.003 & 6.05 & 0.14 & 1 \\
HD 128620/1     & 0.934 & 0.006 & 5.71 & 0.09 & 1 \\
HD 14384        & 1.254 & 0.007 & 3.73 & 0.07 & 1 \\
HD 154676       & 1.220 & 0.006 & 3.76 & 0.10 & 1 \\
HD 179890       & 0.958 & 0.012 & 5.38 & 0.14 & 1 \\
HD 185912       & 1.344 & 0.013 & 3.68 & 0.10 & 1 \\
HD 19457        & 1.347 & 0.008 & 3.46 & 0.07 & 1 \\
HD 206155       & 1.332 & 0.011 & 3.66 & 0.23 & 1 \\
HD 206821       & 1.122 & 0.012 & 4.29 & 0.20 & 1 \\
HD 235444       & 0.932 & 0.007 & 5.37 & 0.18 & 1 \\
HD 287727       & 1.124 & 0.009 & 4.64 & 0.11 & 1 \\
HD 34335        & 1.368 & 0.016 & 3.45 & 0.07 & 1 \\
HD 352179       & 1.408 & 0.009 & 3.13 & 0.12 & 1 \\
HD 37071        & 1.079 & 0.005 & 3.64 & 0.08 & 1 \\
HD 37513        & 1.196 & 0.007 & 3.96 & 0.10 & 1 \\
HD 4161         & 1.352 & 0.009 & 3.09 & 0.19 & 1 \\
HD 62863        & 1.550 & 0.013 & 3.03 & 0.21 & 1 \\
HD 6980         & 1.193 & 0.004 & 3.06 & 0.13 & 1 \\
HD 72257        & 1.146 & 0.006 & 4.14 & 0.13 & 1 \\
HD 7700         & 0.764 & 0.004 & 6.41 & 0.14 & 1 \\
HD 79193        & 1.485 & 0.017 & 3.08 & 0.14 & 1 \\
HD 90242        & 1.219 & 0.007 & 3.86 & 0.06 & 1 \\
HS Aur          & 0.898 & 0.019 & 5.24 & 0.11 & 1 \\
HS Hya          & 1.255 & 0.008 & 3.69 & 0.06 & 1 \\
IT Cas          & 1.332 & 0.009 & 3.22 & 0.09 & 1 \\
NGC188 KR V12   & 1.103 & 0.007 & 3.93 & 0.11 & 1 \\
NGC6791 KR V20  & 0.827 & 0.004 & 6.37 & 0.15 & 1 \\
RT And          & 1.240 & 0.030 & 4.03 & 0.14 & 1 \\
RW Lac          & 0.926 & 0.006 & 4.46 & 0.11 & 1 \\
TYC 3629-740-1  & 0.869 & 0.004 & 5.10 & 0.16 & 1 \\
TYC 3650-959-1  & 1.329 & 0.008 & 3.27 & 0.09 & 1 \\
UV Psc          & 0.983 & 0.008 & 4.58 & 0.11 & 1 \\
UX Men          & 1.235 & 0.006 & 3.80 & 0.10 & 1 \\
V1061 Cyg       & 1.282 & 0.016 & 3.42 & 0.10 & 1 \\
V1143 Cyg       & 1.388 & 0.016 & 3.60 & 0.10 & 1 \\
V1174 Ori       & 1.006 & 0.013 & 5.84 & 0.22 & 1 \\
V505 Per        & 1.272 & 0.007 & 3.66 & 0.07 & 1 \\
V568 Lyr        & 1.074 & 0.008 & 4.19 & 0.11 & 1 \\
V570 Per        & 1.447 & 0.009 & 3.06 & 0.08 & 1 \\
V636 Cen        & 1.052 & 0.005 & 4.66 & 0.09 & 1 \\
VZ Hya          & 1.271 & 0.009 & 3.51 & 0.12 & 1 \\
WZ Oph          & 1.227 & 0.007 & 3.74 & 0.10 & 1 \\
ZZ UMa          & 1.139 & 0.005 & 3.75 & 0.08 & 1 \\
alpha Cen       & 1.105 & 0.007 & 4.33 & 0.06 & 1 \\

G193-027 A & 0.126 & 0.005 & 14.16 & 0.05 & 2 \\
G193-027 B & 0.124 & 0.005 & 14.46 & 0.05 & 2 \\
G250-029 A & 0.350 & 0.005 & 11.07 & 0.03 & 2 \\
G250-029 B & 0.187 & 0.004 & 12.68 & 0.07 & 2 \\
GJ1005 A & 0.179 & 0.002 & 12.70 & 0.01 & 2 \\
GJ1005 B & 0.112 & 0.001 & 15.12 & 0.09 & 2 \\
GJ1081 A & 0.325 & 0.010 & 11.49 & 0.04 & 2 \\
GJ1081 B & 0.205 & 0.007 & 13.16 & 0.09 & 2 \\
GJ1245 A & 0.111 & 0.001 & 15.17 & 0.03 & 2 \\
GJ1245 C & 0.076 & 0.001 & 18.46 & 0.06 & 2 \\
GJ22 A & 0.405 & 0.008 & 10.32 & 0.03 & 2 \\
GJ22 C & 0.157 & 0.003 & 13.40 & 0.10 & 2 \\
GJ234 A & 0.223 & 0.002 & 13.11 & 0.03 & 2 \\
GJ234 B & 0.109 & 0.001 & 16.19 & 0.06 & 2 \\
GJ469 A & 0.332 & 0.007 & 11.69 & 0.03 & 2 \\
GJ469 B & 0.188 & 0.004 & 13.28 & 0.05 & 2 \\
GJ473 A & 0.124 & 0.005 & 15.09 & 0.05 & 2 \\
GJ473 B & 0.113 & 0.005 & 15.08 & 0.05 & 2 \\
GJ54 A & 0.432 & 0.008 & 10.70 & 0.03 & 2 \\
GJ54 B & 0.301 & 0.006 & 11.70 & 0.04 & 2 \\
GJ623 A & 0.379 & 0.007 & 10.77 & 0.03 & 2 \\
GJ623 B & 0.114 & 0.002 & 16.05 & 0.10 & 2 \\
GJ65 A & 0.120 & 0.003 & 15.49 & 0.04 & 2 \\
GJ65 B & 0.117 & 0.003 & 15.94 & 0.05 & 2 \\
GJ748 A & 0.369 & 0.005 & 11.25 & 0.04 & 2 \\
GJ748 B & 0.190 & 0.003 & 13.06 & 0.05 & 2 \\
GJ791.2 A & 0.237 & 0.004 & 13.46 & 0.04 & 2 \\
GJ791.2 B & 0.114 & 0.002 & 16.73 & 0.10 & 2 \\
GJ831 A & 0.270 & 0.004 & 12.66 & 0.02 & 2 \\
GJ831 B & 0.145 & 0.002 & 14.76 & 0.06 & 2 \\

GJ 166 C & 0.177 & 0.029 & 12.68 & 0.03 & 2,3,4 \\
GJ 278 C & 0.599 & 0.005 & 9.01 & 0.33 & 2,1,5 \\
GJ 278 D & 0.599 & 0.005 & 9.01 & 0.33 & 2,1,5 \\
GJ 570 B & 0.586 & 0.007 & 9.45 & 0.05 & 2,6,7 \\
GJ 570 C & 0.390 & 0.005 & 11.09 & 0.17 & 2,7 \\
GJ 630.1 A & 0.231 & 0.001 & 12.72 & 0.02 & 2,8 \\
GJ 630.1 B & 0.214 & 0.001 & 12.86 & 0.03 & 2,8 \\
GJ 747 A & 0.214 & 0.001 & 12.30 & 0.06 & 2,6,9 \\
GJ 747 B & 0.200 & 0.001 & 12.52 & 0.06 & 2,6,9 \\
GJ 860 A & 0.268 & 0.020 & 11.78 & 0.01 & 2,4,10 \\
GJ 860 B & 0.172 & 0.008 & 13.39 & 0.01 & 2,4 \\
GJ 2005 B & 0.079 & 0.020 & 19.24 & 0.07 & 2,11,4 \\
GJ 2005 C & 0.079 & 0.020 & 19.63 & 0.08 & 2,11,4 \\
GU Boo A & 0.616 & 0.006 & 8.60 & 0.17 & 2,12 \\
GU Boo B & 0.600 & 0.006 & 8.89 & 0.18 & 2,12 \\

J045304-0700.4 A & 0.834 & 0.004 & 5.62 & 0.22 & 13,14 \\
J045304-0700.4 B & 0.828 & 0.004 & 5.90 & 0.24 & 13,14 \\
J082552-1622.8 A & 0.729 & 0.004 & 7.44 & 0.35 & 13,14 \\
J082552-1622.8 B & 0.687 & 0.005 & 7.90 & 0.40 & 13,14 \\
J093814-0104.4 A & 0.771 & 0.033 & 7.22 & 0.16 & 13,15 \\
J093814-0104.4 B & 0.768 & 0.021 & 7.23 & 0.16 & 13,15 \\
J212954-5620.1 A & 0.833 & 0.017 & 6.35 & 0.14 & 13,15 \\
J212954-5620.1 B & 0.703 & 0.013 & 7.65 & 0.21 & 13,15 \\
J011328–3821.1 Aa & 0.612 & 0.030 & 9.18 & 0.31 & 13,16 \\
J011328–3821.1 Ab & 0.445 & 0.019 & 10.65 & 0.89 & 13,16 \\

19b-2-01387 A & 0.498 & 0.019 & 10.21 & 0.23 & 13,17* \\
19b-2-01387 B & 0.481 & 0.017 & 10.49 & 0.28 & 13,17* \\
19c-3-01405 A & 0.410 & 0.023 & 11.41 & 0.44 & 13,17* \\
19c-3-01405 B & 0.376 & 0.024 & 11.46 & 0.45 & 13,17* \\
19e-3-08413 A & 0.463 & 0.025 & 10.25 & 0.32 & 13,17* \\
19e-3-08413 B & 0.351 & 0.019 & 11.42 & 0.46 & 13,17* \\
    
\end{longtable}

\vspace{-0.5em}
\noindent
{\footnotesize
\textbf{References.} (1) \citet{torres_accurate_2010}; (2) \citet{benedict_solar_2016}; (3) \citet{holberg_observational_2012}; (4) \citet{henry_optical_1999}; (5) \citet{torres_absolute_2002}; (6) \citet{delfosse_accurate_2000}; (7) \citet{forveille_accurate_1999}; (8) \citet{morales_absolute_2009}; (9) \citet{segransan_accurate_2000}; (10) \citet{tamazian_mk_2006}; (11) \citet{leinert_multiple_2000}; (12) \citet{lopezmorales_gu_2005}; (13) \citet{iglesias-marzoa_refined_2017}; (14) \citet{helminiak_orbital_2011}; (15) \citet{helminiak_orbital_2011-1}; (16) \citet{helminiak_orbital_2012}; (17) \citet{birkby_discovery_2012}; *Values for $M_V$ were determined using given $M_{bol}$ and $T_{eff}$ in \citet{birkby_discovery_2012} and applying bolometric corrections interpolated from empirical data in \citet{pecaut_intrinsic_2013}. 
}

\newpage
\section{Sampled Posterior Distribution}
\begin{figure}[!htbp]
    \centering
    \includegraphics[width=\textwidth]{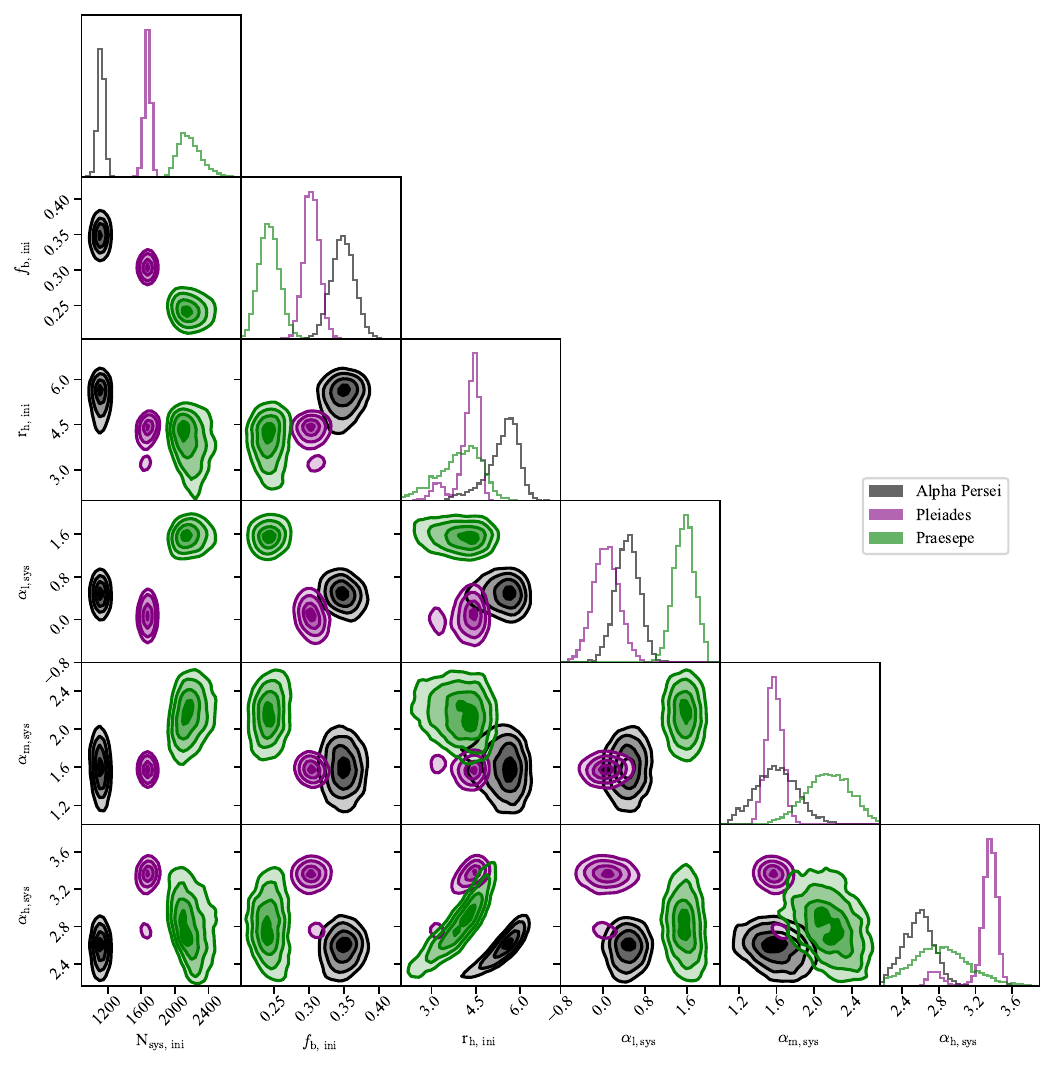}
    \caption{Corner plot of input parameter posterior distribution samples for Alpha Persei, Pleiades and Praesepe. These samples are distributed around the highest likelihood initial parameters for each cluster. We fit for the best combination of initial number of stellar systems ($N_\mathrm{sys,ini}$), initial binary fraction ($f_\mathrm{bin,ini}$), initial 3D half-mass radius ($r_\mathrm{h,ini}$), and the system IMF slopes for the low, intermediate, and high mass ranges ($\alpha_\mathrm{l,sys}$, $\alpha_\mathrm{m,sys}$, $\alpha_\mathrm{h,sys}$).}
    \label{fig:comb_corner}
\end{figure}

\newpage
\section{Raw Uncorrected Cluster Luminosity Functions}

\begin{table}
\centering
\footnotesize
\setlength{\tabcolsep}{4pt}
\renewcommand{\arraystretch}{0.9}
    \begin{tabular}{l S[table-format=3.0] S[table-format=3.0] S[table-format=3.0]}
    \toprule
    M$_G$ Bin Range & {$\alpha$ Per ($N_\mathrm{star}$)} & {Plei ($N_\mathrm{star}$)} & {Prae ($N_\mathrm{star}$)}\\
    \midrule
    $[-2.53,\,-1.87)$ & 1 & 1 & 0 \\
    $[-1.87,\,-1.21)$ & 1 & 2 & 0 \\
    $[-1.21,\,-0.56)$ & 4 & 0 & 0 \\
    $[-0.56,\,0.10)$ & 4 & 6 & 4 \\
    $[0.10,\,0.75)$ & 7 & 3 & 7 \\
    $[0.75,\,1.41)$ & 11 & 11 & 8 \\
    $[1.41,\,2.07)$ & 14 & 10 & 11 \\
    $[2.07,\,2.72)$ & 12 & 20 & 13 \\
    $[2.72,\,3.38)$ & 16 & 13 & 27 \\
    $[3.38,\,4.03)$ & 30 & 22 & 27 \\
    $[4.03,\,4.69)$ & 9 & 29 & 29 \\
    $[4.69,\,5.35)$ & 19 & 35 & 33 \\
    $[5.35,\,6.00)$ & 34 & 37 & 35 \\
    $[6.00,\,6.66)$ & 42 & 39 & 41 \\
    $[6.66,\,7.32)$ & 16 & 43 & 41 \\
    $[7.32,\,7.97)$ & 19 & 54 & 39 \\
    $[7.97,\,8.63)$ & 32 & 63 & 51 \\
    $[8.63,\,9.28)$ & 51 & 54 & 70 \\
    $[9.28,\,9.94)$ & 66 & 101 & 99 \\
    $[9.94,\,10.60)$ & 116 & 129 & 97 \\
    $[10.60,\,11.25)$ & 101 & 209 & 111 \\
    $[11.25,\,11.91)$ & 92 & 190 & 119 \\
    $[11.91,\,12.57)$ & 74 & 125 & 102 \\
    $[12.57,\,13.22)$ & 39 & 82 & 68 \\
    $[13.22,\,13.88)$ & 24 & 54 & 22 \\
    $[13.88,\,14.53)$ & 11 & 27 & 21 \\
    $[14.53,\,15.19]$ & 0 & 16 & 1 \\
    \bottomrule
    \end{tabular}
\caption{Uncorrected observed luminosity functions for Alpha Persei, Pleiades and Praesepe using absolute G band magnitudes M$_G$. Uncertainties follow Poisson statistics, $\sigma = \sqrt{N_\mathrm{star}}$ for each bin.}
\label{tab:lum_func_table}
\end{table}

\begin{figure}[h!]
    \centering
    \includegraphics[width=0.65\textwidth]{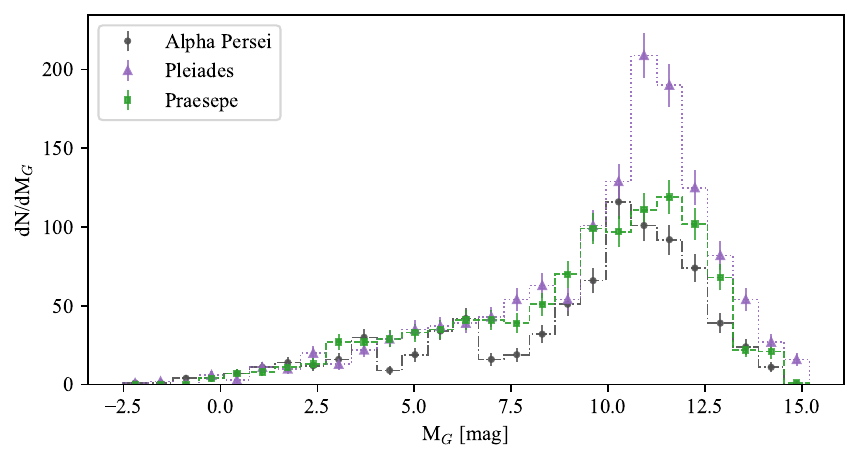}
    \caption{Stellar luminosity functions of Alpha Persei, Pleiades and Praesepe, uncorrected for unresolved binary companions or dynamical evolution. Error bars correspond to $1\sigma=\sqrt{N}$ assuming Poisson statistics. }
    \label{fig:lum_func_plot}
\end{figure}

\end{document}